\documentclass[aps,pra,twocolumn,letterpaper,superscriptaddress,10pt, floatfix]{revtex4-2} 
\usepackage{amssymb,amsthm,amsmath,amsfonts}
\usepackage{graphicx,mathptmx}
\usepackage{upgreek}
\usepackage{textcomp}
\usepackage{amsmath}
\usepackage{array, multirow, bigdelim, makecell, booktabs}
\usepackage{float}

\usepackage[shortlabels]{enumitem}
\usepackage[pdftex,dvipsnames,usenames]{xcolor}
\usepackage[colorlinks=true,urlcolor=blue,citecolor=blue,linkcolor=blue]{hyperref}
\usepackage{gensymb}

\usepackage[dvipsnames]{xcolor}
\usepackage[normalem]{ulem}

\newcommand{\qq}{\mathbf{q}}
\newcommand{\pp}{\mathbf{p}}

\begin{document}

\title{Implementation of space-division multiplexed entanglement-based quantum cryptography over multicore fiber}

\author{Evelyn A. Ortega}
	\email{evelyn.ortega@icfo.eu}
	\affiliation{Institute for Quantum Optics and Quantum Information - IQOQI Vienna, Austrian Academy of Sciences, Boltzmanngasse 3, 1090 Vienna, Austria}
	\affiliation{Vienna Center for Quantum Science and Technology (VCQ), Vienna, Austria}
	\affiliation{Current address: ICFO-Institut  de  Ciencies  Fotoniques,  The  Barcelona  Institute  of Science  and  Technology,  08860  Castelldefels  (Barcelona),  Spain}
	
\author{Jorge Fuenzalida} 
    \email{jorge.fuenzalida@tu-darmstadt.de}
 \affiliation{Technical University of Darmstadt, Department of Physics, Institute for Applied Physics, Schlo{\ss}gartenstraße 7, 64289 Darmstadt, Germany}

\author{Krishna Dovzhik}
	\affiliation{Institute for Quantum Optics and Quantum Information - IQOQI Vienna, Austrian Academy of Sciences, Boltzmanngasse 3, 1090 Vienna, Austria}
	\affiliation{Vienna Center for Quantum Science and Technology (VCQ), Vienna, Austria}
    \affiliation{Current address: Department of Neuroscience and Biomedical Engineering, Aalto School of Science, Aalto University, Espoo, Finland}
	
\author{Rodrigo F. Shiozaki}
    \affiliation{Departamento de F\'isica, Universidade Federal de S\~{a}o Carlos, Rodovia Washington Lu\'is, km 235—SP-310, 13565-905 S\~{a}o Carlos, SP, Brazil}

\author{Juan Carlos Alvarado-Zacarias}
    \affiliation{CREOL, The University of Central Florida, Orlando, Florida 32816, USA}
    
\author{Rodrigo Amezcua-Correa}
    \affiliation{CREOL, The University of Central Florida, Orlando, Florida 32816, USA}
    
\author{Martin Bohmann}
    \affiliation{Institute for Quantum Optics and Quantum Information - IQOQI Vienna, Austrian Academy of Sciences, Boltzmanngasse 3, 1090 Vienna, Austria}
	\affiliation{Vienna Center for Quantum Science and Technology (VCQ), Vienna, Austria}
 \affiliation{Current address: Quantum Technology Laboratories GmbH, Clemens-Holzmeister-Straße 6/6, 1100 Vienna, Austria}
	
\author{Sören Wengerowsky}
    \affiliation{Institute for Quantum Optics and Quantum Information - IQOQI Vienna, Austrian Academy of Sciences, Boltzmanngasse 3, 1090 Vienna, Austria}
	\affiliation{Vienna Center for Quantum Science and Technology (VCQ), Vienna, Austria}
	\affiliation{Current address: ICFO-Institut  de  Ciencies  Fotoniques,  The  Barcelona  Institute  of Science  and  Technology,  08860  Castelldefels  (Barcelona),  Spain}
	
\author{Rupert Ursin}
	\affiliation{Institute for Quantum Optics and Quantum Information - IQOQI Vienna, Austrian Academy of Sciences, Boltzmanngasse 3, 1090 Vienna, Austria}
	\affiliation{Vienna Center for Quantum Science and Technology (VCQ), Vienna, Austria}
 \affiliation{Current address: Quantum Technology Laboratories GmbH, Clemens-Holzmeister-Straße 6/6, 1100 Vienna, Austria}

\begin{abstract}
Quantum communication implementations require efficient and reliable quantum channels.
Optical fibers have proven to be an ideal candidate for distributing quantum states.
Thus, today's efforts address overcoming issues towards high data transmission and long-distance implementations.
Here, we experimentally demonstrate the secret key rate enhancement via space-division multiplexing using a multicore fiber.
Our multiplexing technique exploits the momentum correlation of photon pairs generated by spontaneous parametric down-conversion.
We distributed polarization-entangled photon pairs into opposite cores within a 19-core multicore fiber. 
We estimated the secret key rates in a configuration with 6 and 12 cores from the entanglement visibility after transmission through $411$~m long multicore fiber.
\end{abstract}

\date{\today}
\maketitle

\section{Introduction}

Quantum key distribution (QKD) is an established quantum communication method that allows generation of identical secret keys between two remotely separated parties with unconditional security~\cite{pirandola2020advances, RevModPhys.92.025002}.
The communication protocols to share the key can be based on entangled states~\cite{ursin2007entanglement,cabrejo2022ghz} or weak coherent states~\cite{lo2005decoy,wang2022twin}.
This key needs to be shared by trusted channels such as free-space links on ground~\cite{Scheidl_2009,krvzivc2022metropolitan}, ground-to-space~\cite{liao2017satellite,yin2020entanglement} and optical fiber links~\cite{boaron2018quantum, Shi2020,chen2022quantum}.
Over the years, improvements in optical fiber fabrication have enabled access to new fibers, which offer flexibility, reliability, and high fidelity in data transmission~\cite{Winzer:18}.
This has allowed implementing quantum key distribution schemes based on single-photon sources over single-mode fibers (SMF)~\cite{ribezzo2022deploying}, few-mode fibers~\cite{Wang:20,alarcon2021few}, multimode fibers~\cite{zhou2021high,amitonova2020quantum}, vortex fibers~\cite{sit2018quantum}, and multicore fibers (MCF)~\cite{Dynes:16, bacco2017space}.
Recently, schemes based on entangled-photon states have been shown to generate a positive key rate over a long-distance fiber link~\cite{Wengerowsky2020,neumann2022quapital}.
These schemes represent a challenge due to high propagation losses in fiber, polarization mode dispersion, and stabilization of the state in the propagation.

In order to implement quantum communication schemes over long fiber links, it is crucial to develop techniques to increase channel transmission capacity.
One of the most widely used techniques for that is multiplexing, by transmitting more than one signal through the same communication channel.
This has been implemented based on the wavelength (WDM) degree of freedom to increase the secure key rate~\cite{neumann2022experimental, brambila2023ultrabright} and carry out novel entanglement-based quantum networks~\cite{wengerowsky2018entanglement,joshi2020trusted}.
But it is also crucial to explore new domains such as space-division multiplexing (SDM) to overcome WDM's capacity~\cite{richardson2013, Puttnam:21,xavier2020quantum}.
In particular, SDM technology over multicore fiber (MCF) has captured attention due to the low crosstalk between cores.
MCF has been employed as a quantum communication link to increase the secret key rate~\cite{Bacco2019}, implementation of QKD schemes included integrated into silicon photonic chips~\cite{bacco2017space}, simultaneous transmission of classical and quantum signals over a single MCF~\cite {Dynes:16}, and generation of high-dimensional states~\cite{canas2017decoy}.
Furthermore, the multiple fibers available over the same fiber link show an excellent performance to distribute entangled states based on their spatial correlation~\cite{ortega2021experimental,achatz2022simultaneous}.

In this work, we experimentally demonstrate generation of parallel and independent quantum keys based on SDM over MCF using a single entangled photon source.
The multiplexing is performed through the momentum correlation of the photon pairs, whereas the secure key rate is extracted from the polarization measurement outcomes by each correlated channel separately.
In particular, we report the key rate associated with the 19-core configurations into the MCF, corresponding to a SDM scheme over six and twelve cores.
We extrapolate the achievable key rate for the case of a fast-switching QKD setup as a function of MCF length to explore its applicability.
In addition, we demonstrate a stable and low quantum bit error rate (QBER) below the threshold limit, along with a positive key rate for one pair of opposite cores.
For at least 24 hours of continuous operation, polarization changes were observed without any active stabilization technique.
Thus, our results show the first step towards the performance of SDM on entanglement-based quantum communication protocol and the development of efficient quantum communication distribution.

\section{Results}
\subsection{Experimental setup}

\begin{figure*}[ht]
\centering
\includegraphics[width=1\textwidth]{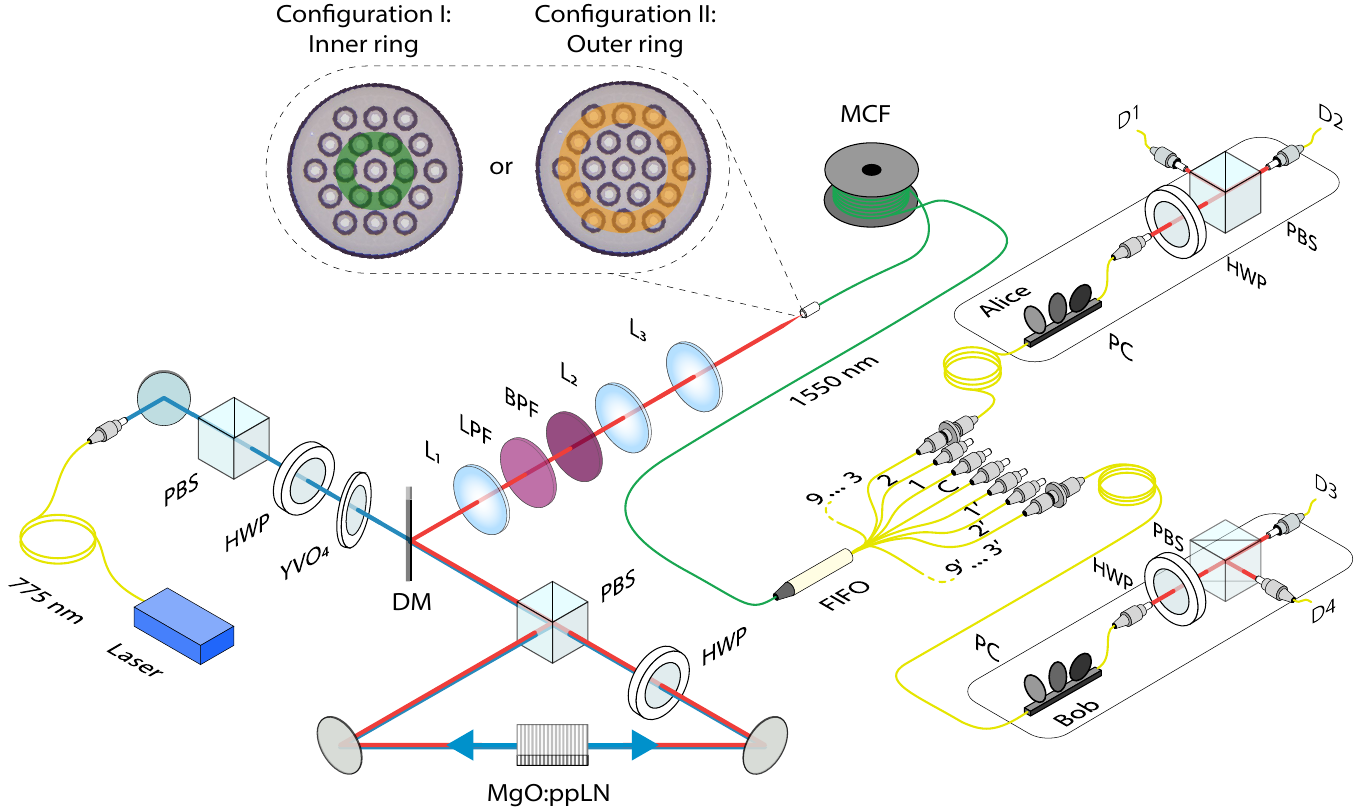}
\caption{Experimental setup.
A MgO:ppLN crystal is pumped in a Sagnac configuration by a continuous wave laser at $775$~nm (depicted as a blue beam) to create polarization-entangled photon pairs with the center wavelength of $1550$~nm (depicted as a red beam). 
The signal and idler photons were separated from the pump with a dichroic mirror (DM) and long-pass filter (LPF) to be distributed through opposite cores of the MCF employing the spatial correlations of the SPDC photons.
These correlations are imaged onto the fiber end-face by a combination of lenses $\text L_{1}$, $\text L_{2}$, and $\text L_{3}$. 
The SPDC photons are emitted in a cone, whose size is adjusted to fit hexagonal rings of cores (either inner or outer) by controlling the crystal temperature~\cite{ortega2022spatial}.
The focal lengths of the lenses are $\text f_{1}=200$~mm, $\text f_{2}=150$~mm and $\text f_{3}= 4.51$~mm.
The 19 cores are split into individual SMF via a fan-in/fan-out (FIFO) device. 
Each pair of opposite cores are connected to a  polarization analysis module (Alice and Bob) consisting of a half-wave plate (HWP) in front of a polarizing beam splitter (PBS).
D1 through D4 refer to detectors (superconducting nanowire single-photon detectors, SNSPD) used to detect the photons.
Abbreviations: YVO$_{4}$: yttrium orthovanadate plate, BPF: 1550$\pm$3~nm band-pass interference filter, PC: polarization controllers.}
\label{fig:setup}
\end{figure*}

Our experimental setup is depicted in Fig.~\ref{fig:setup}.
We generated photon pairs at telecommunication wavelength from a type-0, non-collinear, quasi-phase-matched spontaneous parametric down-conversion (SPDC) process.
By bidirectionally pumping a magnesium oxide doped periodically poled lithium niobate (MgO:ppLN) crystal in a Sagnac configuration, the SPDC source produces polarization-entangled photon pairs in the Bell state,
\begin{equation}
\left\vert \Phi^{+}\right\rangle = \frac{1}{\sqrt{2}}(\left\vert H_{s},H_{i}\right\rangle+\left\vert V_{\mathrm{s}},V_{i}\right\rangle),
\label{eq:state2}
\end{equation}
where H (V) indicates horizontal (vertical) polarization for signal (s) and idler (i) photons.
Exploiting momentum conservation in the SPDC process, signal and idler photons were collected onto the MCF's opposite cores, which were located at the far-field plane of the crystal \cite{ortega2021experimental},
\begin{align}
    |\boldsymbol{\Phi}\rangle=\sum_{m=1}^N g_m|c_m,c_m^{\prime}\rangle_{\mathrm{cores}},
\end{align}
where $c_m$ and $c_m^{\prime}$ corresponds to the opposite cores, $g_m$ are normalized probability amplitudes, $\sum_{m=1}^N |g_m|^2=1 $, and $N$ is the number of opposite cores.
The far-field plane contains the transverse momentum of the signal ($\pp_s=\hbar \qq_s$) and idler ($\pp_i=\hbar \qq_i$), where $\qq$ represents the transverse wave vector~\cite{WALBORN201087, Achatz_2022}.
This plane is accessed by performing a Fourier transform of the photons source's plane with a configuration of lenses as shown in Fig.~\ref{fig:setup}.

The MCF selected for the experiment is $411$~m in length, with 19 cores distributed in a central core and two hexagonal rings of cores, which we refer to as the inner ring (6 cores) and the outer ring (12 cores).
This MCF presents very low cross-talk between adjacent cores thanks to a trench-assisted structure around each core~\cite{trenchMCF2011}.
Additionally, each of these cores ends into an independent SMF via a fan-in/fan-out (FIFO) device~\cite{MCF-JC}.
In order to couple the polarization-entangled photon pairs into opposite cores of each MCF ring independently, the emission angle of the SPDC photons was adjusted via temperature control of the nonlinear crystal~\cite{ortega2022spatial}.
Thus, we can reliably and quickly switch between a different number of channels (cores), extending the generation of the simultaneous secure keys without changing optical components in the SPDC source.

To estimate the quantum key rate from each pair of opposite cores (one at a time), entangled photons were distributed into the polarization analysis modules to Alice and Bob through the SMF from the FIFO device.
In both modules, the correlated photons were detected by a superconducting nanowire single-photon detector (SNSPD) operated at efficiencies of $\sim80\%$ with dark count rates of $\sim100$ counts per second.
A time-tagging module recorded the detection events of each pair of opposite cores.
Correlated photon pairs were identified through coincidence counts with the help of the temporal cross-correlation functions within $300$~ps time-window.

\subsection{Estimation of the secure key rate over MCF}

We calculated the secure key rate based on the count rates for each pair of opposite cores~\cite{QKD_Ma,neumann2021} and theoretical prediction of its propagation, assuming an MCF average loss of $0.2$~dB/km per core.
A series of polarization measurements in two mutually unbiased bases allowed us to estimate the secure key rates on opposite cores in the Horizontal-Vertical (HV)- basis and in the Diagonal-Antidiagonal (DA) basis.
Since our system lacked the ability to have fast and random basis choices, we estimated the key rate from the polarization visibility measurement that we would have observed if we had a passive basis choice on a beam splitter and four detectors, two for each basis.
Instead, we used two motor-driven rotation stages on Alice and Bob to set the half-wave plates between $0^{\circ}$~(H/V basis) and $22.5^{\circ}$~(D/A basis).
They were followed by a polarizing beam-splitter (PBS) with a fiber-coupler in order to connect the transmitted and reflected ports to separate detectors, respectively.
For each pair of opposite cores, three pairs for the inner ring and six pairs for the outer ring, the data was accumulated for $60\,$~s.

\begin{figure}[h!]
\centering
\includegraphics[width=1\columnwidth]{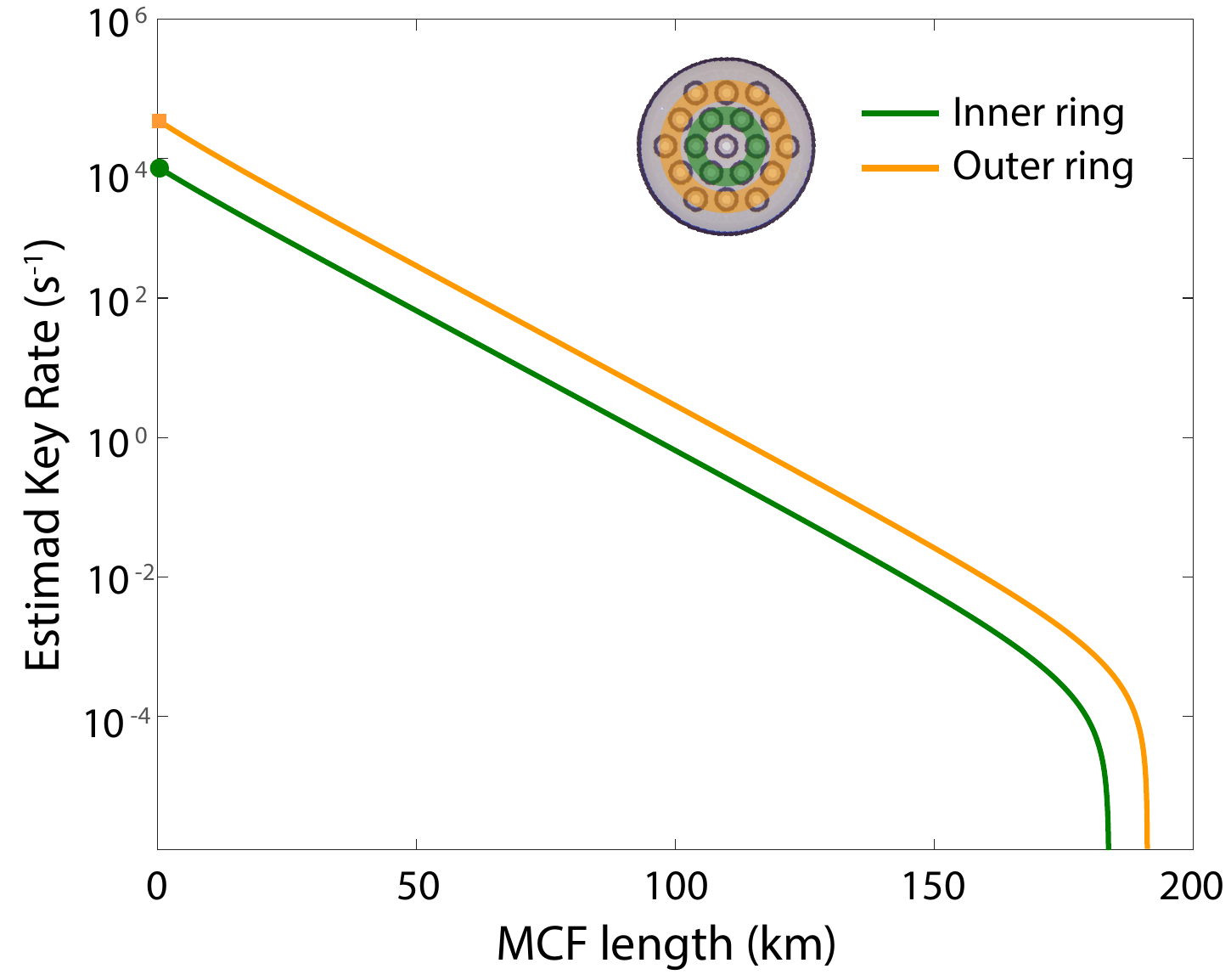}
\caption{Estimated secret key rate as a function of the MCF length. 
The green circle represents the overall secret key rate of the inner ring obtained after a propagation distance of $411$~m in a MCF. 
A total key rate of $7.3$~kbit/s was estimated based on the count rates and visibilities measured for a SDM scheme over 6 cores, i.e., three pairs of opposite cores.
The orange square corresponds to the overall secret key rate of the outer ring obtained after the same propagation length.
In this case, we increased the total key rate up to $34.5$~kbit/s through an SDM scheme over 12 cores, i.e., six pairs of cores.}
\label{fig:skr}
\end{figure}

Figure~\ref{fig:skr} shows the estimated key rate produced for the inner and outer rings corresponding to green and orange dots, respectively and a theoretical prediction of how much key rate would be observed under the same conditions as a function of channel length represented by the solid lines.
For the three pairs of cores of the inner ring, the average coincidence rate was $2287$~counts per second on the H/V basis and $2275$~counts per second on the D/A basis.
For the six pairs of cores of the outer ring, the average coincidence rate was $7832$~counts per second on the H/V basis and $7770$~counts per second on the D/A basis.
Based on these coincidence count rates, we estimated an overall secure key rate of $7.3$~kbit/s for the inner ring and $34.5$~kbit/s for the outer ring.
Therefore, using one entanglement photon source and multiplexing in different spatial modes, we achieve up to $4.7$~times higher key rates in the outer ring than in the inner ring.
One would expect a key rate two times higher in the outer ring than the inner ring due to the number of cores, the same number of photon pairs generated, and the identical detection modules.
However, we can observe a higher average coincidence rate in the outer ring, which may be due to the different crystal temperatures of both configurations, which directly affect SPDC photon emission and spatial correlations~\cite{ortega2022spatial}.
The spatial correlation increases for a lower crystal temperature, helping to increase coincidence counts of the polarization-entangled photon pairs and reducing accidental coincidence counts.
Indeed, our experimental implementation turns out to have different loss levels for the inner ring of $40.06$~dB and the outer ring of $35.48$~dB.
From the theoretical prediction shown in Fig.~\ref{fig:skr}, the expected key rates that can be attested to a MCF link have a limiting distance of $180$~km and $190$~km for the inner and outer rings, respectively.
Up to which the key rate would be positive in the infinite key limit.

\begin{figure}[h!]
\centering
\includegraphics[width=1\columnwidth]{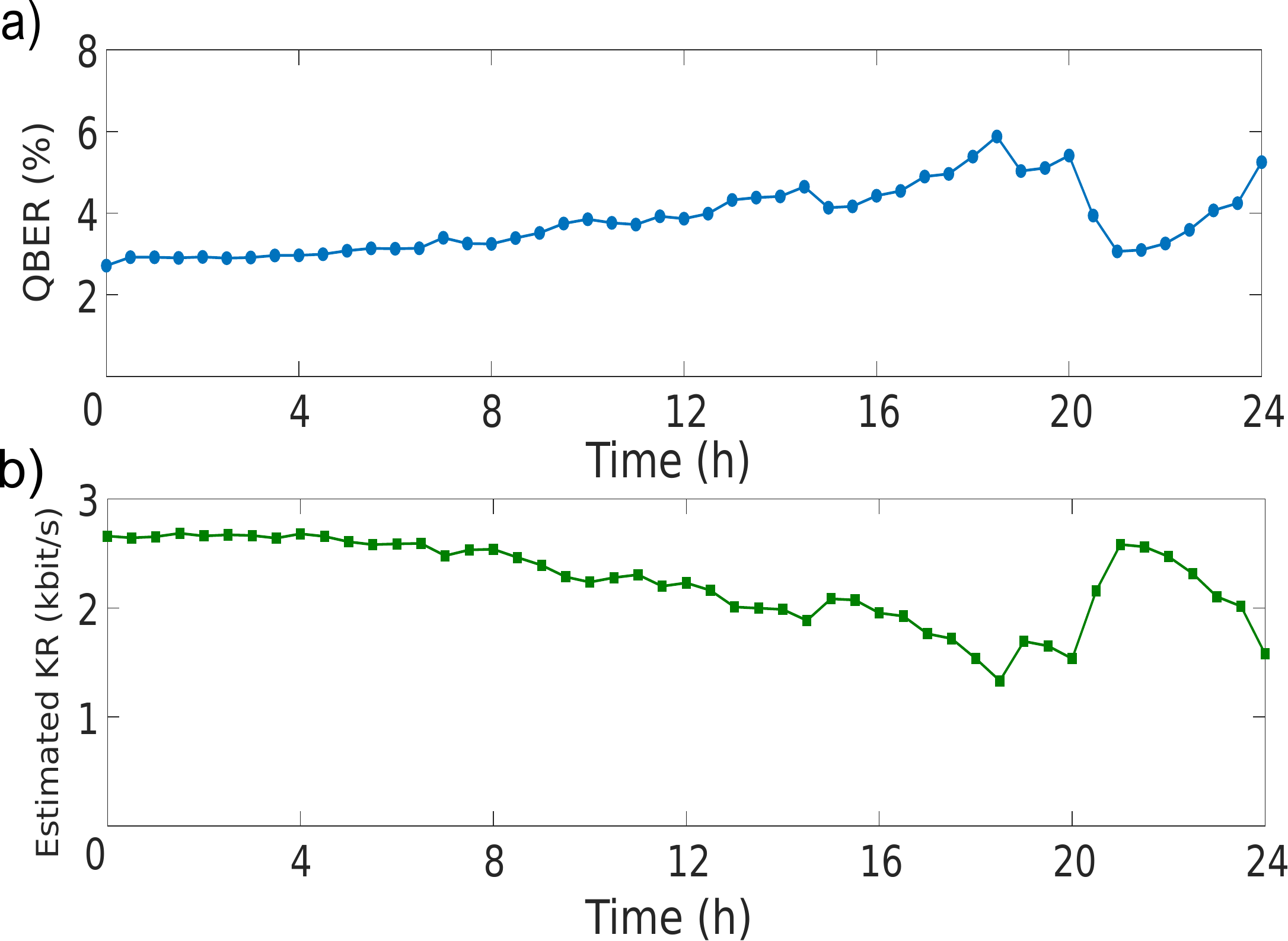}
\caption{Estimated a) QBER and b) key rate (KR) over $24$~hours for one pair of opposite cores from the inner ring. 
Each data point was calculated from the polarization visibility measurement in the HV and DA bases over $60$~s, changing bases every $30$~min.}
\label{fig:QBER}
\end{figure}

Additionally, we performed a study on the system stability by a long-term measurement of coincidence counts in the H/V and D/A bases for one pair of opposite cores from the inner ring throughout $24$~hours.
The coincidence counts were collected every $30$~min, alternating the measurement basis for periods of $60$~s.
As evidenced in Fig.~\ref{fig:QBER} (a), QBER (blue dots) remains stable around $3\%$ for the whole measurement without any active stabilization. 
It is worth mentioning that a QBER below $11\%$ ensures a positive secure key rate~\cite{neumann2021}.
In the other cores of the ring, we estimated the QBER over $60\,$~s, obtaining in all the combinations QBERs lower than $8.2\%$. 
The lowest QBER that can be reached in the inner ring was $3.02\%$ on a H/V basis and $2.42\%$ on a D/A basis. 
Furthermore, these minima in the outer ring were $2.65\%$ on a H/V basis and $2.72\%$ on a  D/A basis.
While Fig.~\ref{fig:QBER} (b) shows the estimated key rate around $2.3$~kbit/s (green dots) based on the coincidence rate, considering the error correction efficiency of 1.2 \cite{Elkouss2009}.

Hence, employing intrinsic correlations in an SPDC source, it is feasible to implement a deterministic spatial multiplexing scheme for quantum communication by extracting independent secure keys of entangled photon pairs.
Note that in our SDM implementation, increasing the number of cores is always possible by adjusting the crystal temperature, which optimizes the momentum correlation and allows for avoiding crosstalk.
Up to now, MCFs with 37 cores have been used to implement experimental QKD by employing a classical signal~\cite{Bacco2019}.
This approach adds a third hexagonal ring of 18 cores around our cores configuration (19 cores)~\cite{sasaki2017single}.
With such configuration, the third ring would allow us to connect nine additional core pairs, which would allow obtaining a secret key rate at least 1.5 times larger than the key rate of the outer ring, considering the same loss level.
\newpage
\section{Discussion}

We have experimentally shown a space-division multiplexing technique to enhance the total key rate and, therefore, with promising applications in quantum communication.
Our SDM scheme uses a 19-core MCF wherein we distributed polarization entangled-photon pairs between 3 pairs of users (6 cores; inner ring) and 6 pairs of users (12 cores; outer ring).
The switch between these two configurations was done by tuning the crystal temperature. 
The secure key rate was computed by polarization visibility measurements between independent pairs of cores.
We obtained a coincidence rate for the 6 and 12 cores in the order of $2000$ and $7000$ counts per second, respectively.
With these counts, we computed a key rate of $7.3$~kbit/s ($34.5$~kbit/s) in the inner (outer) ring.
In a second analysis, we verified the estimated QBER and key rate for a pair of cores over 10 hours.
Although no active polarization stabilization techniques were employed, the values remained stable around $2.2$~kbit/s and a QBER of $3.8\%$. 

Our results show a feasible implementation where a 19-core MCF can generate parallel key rates in independent cores that can be combined, distributed individually, or grouped for a multi-user quantum network.
This scheme can be easily adapted to a higher number of cores or combined with other degrees of freedom for enhancing the key rate through a MCF~\cite{achatz2022simultaneous}.
However, our experimental implementation can be improved in several aspects, such as the coupling efficiency of the lens system to implement the far-field plane, which can be overcome by employing, e.g., a microlens array~\cite{Dietrich:17}.
Also, the parameters of the SPDC source, such as the length of the crystal or pump beam waist, could be optimized to increase the spatial entanglement.
An enhancement in the coupling efficiency could allow the integration of such an SDM scheme into existing telecommunication infrastructures operating at telecommunication wavelength.
Therefore, our work opens the possibility of employing MCF for various quantum communication applications.


\section*{Acknowledgments}
We acknowledge funding by the European Union’s Horizon 2020 programme grant agreement No.857156 (OpenQKD) and the Austrian Academy of Sciences.
J.C.A.-Z. and R.A.-C. acknowledge support US. Army Research Office W911NF1710553 and National Science Foundation ECCS-1711230.
E.A.O. acknowledges ANID for the financial support (Becas de doctorado en el extranjero “Becas Chile”/2016 – No. 72170402).

\section{Methods} 
\label{sec:method}

\subsection{Entangled photon pair source}

The SPDC-source is based on $40$~mm long Magnesium Oxide doped periodically poled Lithium Niobate (MgO:ppLN) crystal with a poling period of $19.2$~$\upmu$m placed in a Sagnac interferometer~\cite{kim2006}.
The nonlinear crystal was pumped by a continuous-wave laser at the center wavelength $775$~nm.
In the down-conversion process, one pump photon is converted with low probability to a pair of signal and idler photons at telecommunication wavelength around $1550$~nm with the same polarization as the pump photon.

The far-field plane of the crystal was accessed by the lens $\text L_{1}$, where the transverse detection positions of the photon pairs are spatially distributed in opposite directions.
So the photon pairs in the far-field plane were coupled into the MCF by the 4f-imaging system consisting of the lenses $\text L_{2}$ and $\text L_{3}$.
This demagnified image in the far-field plane ensured that the emission cone diameter covered all cores of the corresponding ring.
Additionally, for adjusting the SPDC emission cone diameter to the inner and outer ring, the crystal temperature is set at $82.5^{\circ}$C and $82^{\circ}$C, respectively. 

\subsection{Estimation of key rate}

For each pair of opposite cores, the secret key rate has been estimated based on the assumption that the setup used a fast and random basis choice for each photon~\cite{QKD_Ma,neumann2021}.
Considering $H_{2}$ the binary Shannon entropy function defined as
\begin{equation}
\label{eq:shannonentropy}
   H_{2}\left(x\right)=-x\log_{2}\left(x\right)-\left(1-x\right)\log_{2}\left(1-x\right).
\end{equation}
The secret key rate is given by:
\begin{equation}
\label{eq:SKR}
\begin{split}
    R& = \frac{1}{2}C_{HV}\left(1-\left(1+f\right)H_{2}\left(Q_{HV}\right)\right)\\
                 &\quad  +\frac{1}{2}C_{DA}\left(1-(1+f)H_{2}\left(Q_{DA}\right)\right),
\end{split}
\end{equation}
where $C_{HV}$ and $C_{DA}$ are the total number of coincidence counts in the H/V and D/A bases, respectively.
$Q_{HV}$ and $Q_{DA}$ are the QBER in the H/V and D/A bases, respectively, which can be calculated from the coincidence counts at the given polarization combination over $60$~s via
\begin{equation}
\label{eq:QBER}
\centering
\begin{split}
  Q_{HV}&= \frac{\textup{error~counts}}{\textup{total~counts}} \\
        &= \frac{1-V_{HV}}{2}.
\end{split}
\end{equation}
Where $V_{HV}$ corresponds to the visibility in H/V basis,
\begin{equation}
    V_{HV} =\frac{C_{HH}+C_{VV}-C_{HV}-C_{VH}}{C_{HH}+C_{VV}+C_{HV}+C_{VH}}
\label{eq:polvis}
\end{equation}
with $C_{s_{1},s_{2}}$ is the number of coincidence counts for each polarization setting.
The QBER on DA basis is calculated analogously.  


\bibliography{biblio}

\begin{thebibliography}{46}%
\makeatletter
\providecommand \@ifxundefined [1]{%
 \@ifx{#1\undefined}
}%
\providecommand \@ifnum [1]{%
 \ifnum #1\expandafter \@firstoftwo
 \else \expandafter \@secondoftwo
 \fi
}%
\providecommand \@ifx [1]{%
 \ifx #1\expandafter \@firstoftwo
 \else \expandafter \@secondoftwo
 \fi
}%
\providecommand \natexlab [1]{#1}%
\providecommand \enquote  [1]{``#1''}%
\providecommand \bibnamefont  [1]{#1}%
\providecommand \bibfnamefont [1]{#1}%
\providecommand \citenamefont [1]{#1}%
\providecommand \href@noop [0]{\@secondoftwo}%
\providecommand \href [0]{\begingroup \@sanitize@url \@href}%
\providecommand \@href[1]{\@@startlink{#1}\@@href}%
\providecommand \@@href[1]{\endgroup#1\@@endlink}%
\providecommand \@sanitize@url [0]{\catcode `\\12\catcode `\$12\catcode
  `\&12\catcode `\#12\catcode `\^12\catcode `\_12\catcode `\%12\relax}%
\providecommand \@@startlink[1]{}%
\providecommand \@@endlink[0]{}%
\providecommand \url  [0]{\begingroup\@sanitize@url \@url }%
\providecommand \@url [1]{\endgroup\@href {#1}{\urlprefix }}%
\providecommand \urlprefix  [0]{URL }%
\providecommand \Eprint [0]{\href }%
\providecommand \doibase [0]{https://doi.org/}%
\providecommand \selectlanguage [0]{\@gobble}%
\providecommand \bibinfo  [0]{\@secondoftwo}%
\providecommand \bibfield  [0]{\@secondoftwo}%
\providecommand \translation [1]{[#1]}%
\providecommand \BibitemOpen [0]{}%
\providecommand \bibitemStop [0]{}%
\providecommand \bibitemNoStop [0]{.\EOS\space}%
\providecommand \EOS [0]{\spacefactor3000\relax}%
\providecommand \BibitemShut  [1]{\csname bibitem#1\endcsname}%
\let\auto@bib@innerbib\@empty
\bibitem [{\citenamefont {Pirandola}\ \emph {et~al.}(2020)\citenamefont
  {Pirandola}, \citenamefont {Andersen}, \citenamefont {Banchi}, \citenamefont
  {Berta}, \citenamefont {Bunandar}, \citenamefont {Colbeck}, \citenamefont
  {Englund}, \citenamefont {Gehring}, \citenamefont {Lupo}, \citenamefont
  {Ottaviani} \emph {et~al.}}]{pirandola2020advances}%
  \BibitemOpen
  \bibfield  {author} {\bibinfo {author} {\bibfnamefont {S.}~\bibnamefont
  {Pirandola}}, \bibinfo {author} {\bibfnamefont {U.~L.}\ \bibnamefont
  {Andersen}}, \bibinfo {author} {\bibfnamefont {L.}~\bibnamefont {Banchi}},
  \bibinfo {author} {\bibfnamefont {M.}~\bibnamefont {Berta}}, \bibinfo
  {author} {\bibfnamefont {D.}~\bibnamefont {Bunandar}}, \bibinfo {author}
  {\bibfnamefont {R.}~\bibnamefont {Colbeck}}, \bibinfo {author} {\bibfnamefont
  {D.}~\bibnamefont {Englund}}, \bibinfo {author} {\bibfnamefont
  {T.}~\bibnamefont {Gehring}}, \bibinfo {author} {\bibfnamefont
  {C.}~\bibnamefont {Lupo}}, \bibinfo {author} {\bibfnamefont {C.}~\bibnamefont
  {Ottaviani}}, \emph {et~al.},\ }\bibfield  {title} {\bibinfo {title}
  {Advances in quantum cryptography},\ }\href
  {https://doi.org/https://doi.org/10.1364/AOP.361502} {\bibfield  {journal}
  {\bibinfo  {journal} {Adv. Opt. Photon.}\ }\textbf {\bibinfo {volume} {12}},\
  \bibinfo {pages} {1012} (\bibinfo {year} {2020})}\BibitemShut {NoStop}%
\bibitem [{\citenamefont {Xu}\ \emph {et~al.}(2020)\citenamefont {Xu},
  \citenamefont {Ma}, \citenamefont {Zhang}, \citenamefont {Lo},\ and\
  \citenamefont {Pan}}]{RevModPhys.92.025002}%
  \BibitemOpen
  \bibfield  {author} {\bibinfo {author} {\bibfnamefont {F.}~\bibnamefont
  {Xu}}, \bibinfo {author} {\bibfnamefont {X.}~\bibnamefont {Ma}}, \bibinfo
  {author} {\bibfnamefont {Q.}~\bibnamefont {Zhang}}, \bibinfo {author}
  {\bibfnamefont {H.-K.}\ \bibnamefont {Lo}},\ and\ \bibinfo {author}
  {\bibfnamefont {J.-W.}\ \bibnamefont {Pan}},\ }\bibfield  {title} {\bibinfo
  {title} {Secure quantum key distribution with realistic devices},\ }\href
  {https://doi.org/10.1103/RevModPhys.92.025002} {\bibfield  {journal}
  {\bibinfo  {journal} {Rev. Mod. Phys.}\ }\textbf {\bibinfo {volume} {92}},\
  \bibinfo {pages} {025002} (\bibinfo {year} {2020})}\BibitemShut {NoStop}%
\bibitem [{\citenamefont {Ursin}\ \emph {et~al.}(2007)\citenamefont {Ursin},
  \citenamefont {Tiefenbacher}, \citenamefont {Schmitt-Manderbach},
  \citenamefont {Weier}, \citenamefont {Scheidl}, \citenamefont {Lindenthal},
  \citenamefont {Blauensteiner}, \citenamefont {Jennewein}, \citenamefont
  {Perdigues}, \citenamefont {Trojek} \emph {et~al.}}]{ursin2007entanglement}%
  \BibitemOpen
  \bibfield  {author} {\bibinfo {author} {\bibfnamefont {R.}~\bibnamefont
  {Ursin}}, \bibinfo {author} {\bibfnamefont {F.}~\bibnamefont {Tiefenbacher}},
  \bibinfo {author} {\bibfnamefont {T.}~\bibnamefont {Schmitt-Manderbach}},
  \bibinfo {author} {\bibfnamefont {H.}~\bibnamefont {Weier}}, \bibinfo
  {author} {\bibfnamefont {T.}~\bibnamefont {Scheidl}}, \bibinfo {author}
  {\bibfnamefont {M.}~\bibnamefont {Lindenthal}}, \bibinfo {author}
  {\bibfnamefont {B.}~\bibnamefont {Blauensteiner}}, \bibinfo {author}
  {\bibfnamefont {T.}~\bibnamefont {Jennewein}}, \bibinfo {author}
  {\bibfnamefont {J.}~\bibnamefont {Perdigues}}, \bibinfo {author}
  {\bibfnamefont {P.}~\bibnamefont {Trojek}}, \emph {et~al.},\ }\bibfield
  {title} {\bibinfo {title} {Entanglement-based quantum communication over 144
  km},\ }\href {https://doi.org/10.1038/nphys629} {\bibfield  {journal}
  {\bibinfo  {journal} {Nature Phys.}\ }\textbf {\bibinfo {volume} {3}},\
  \bibinfo {pages} {481} (\bibinfo {year} {2007})}\BibitemShut {NoStop}%
\bibitem [{\citenamefont {Cabrejo-Ponce}\ \emph {et~al.}(2022)\citenamefont
  {Cabrejo-Ponce}, \citenamefont {Spiess}, \citenamefont {Muniz}, \citenamefont
  {Ancsin},\ and\ \citenamefont {Steinlechner}}]{cabrejo2022ghz}%
  \BibitemOpen
  \bibfield  {author} {\bibinfo {author} {\bibfnamefont {M.}~\bibnamefont
  {Cabrejo-Ponce}}, \bibinfo {author} {\bibfnamefont {C.}~\bibnamefont
  {Spiess}}, \bibinfo {author} {\bibfnamefont {A.~L.~M.}\ \bibnamefont
  {Muniz}}, \bibinfo {author} {\bibfnamefont {P.}~\bibnamefont {Ancsin}},\ and\
  \bibinfo {author} {\bibfnamefont {F.}~\bibnamefont {Steinlechner}},\
  }\bibfield  {title} {\bibinfo {title} {Ghz-pulsed source of entangled photons
  for reconfigurable quantum networks},\ }\href
  {https://iopscience.iop.org/article/10.1088/2058-9565/ac86f0/meta} {\bibfield
   {journal} {\bibinfo  {journal} {Quantum Sci. Technol.}\ }\textbf {\bibinfo
  {volume} {7}},\ \bibinfo {pages} {045022} (\bibinfo {year}
  {2022})}\BibitemShut {NoStop}%
\bibitem [{\citenamefont {Lo}\ \emph {et~al.}(2005)\citenamefont {Lo},
  \citenamefont {Ma},\ and\ \citenamefont {Chen}}]{lo2005decoy}%
  \BibitemOpen
  \bibfield  {author} {\bibinfo {author} {\bibfnamefont {H.-K.}\ \bibnamefont
  {Lo}}, \bibinfo {author} {\bibfnamefont {X.}~\bibnamefont {Ma}},\ and\
  \bibinfo {author} {\bibfnamefont {K.}~\bibnamefont {Chen}},\ }\bibfield
  {title} {\bibinfo {title} {Decoy state quantum key distribution},\ }\href
  {https://doi.org/10.1103/PhysRevLett.94.230504} {\bibfield  {journal}
  {\bibinfo  {journal} {Phys. Rev. Lett.}\ }\textbf {\bibinfo {volume} {94}},\
  \bibinfo {pages} {230504} (\bibinfo {year} {2005})}\BibitemShut {NoStop}%
\bibitem [{\citenamefont {Wang}\ \emph {et~al.}(2022)\citenamefont {Wang},
  \citenamefont {Yin}, \citenamefont {He}, \citenamefont {Chen}, \citenamefont
  {Wang}, \citenamefont {Ye}, \citenamefont {Zhou}, \citenamefont {Fan-Yuan},
  \citenamefont {Wang}, \citenamefont {Zhu} \emph {et~al.}}]{wang2022twin}%
  \BibitemOpen
  \bibfield  {author} {\bibinfo {author} {\bibfnamefont {S.}~\bibnamefont
  {Wang}}, \bibinfo {author} {\bibfnamefont {Z.-Q.}\ \bibnamefont {Yin}},
  \bibinfo {author} {\bibfnamefont {D.-Y.}\ \bibnamefont {He}}, \bibinfo
  {author} {\bibfnamefont {W.}~\bibnamefont {Chen}}, \bibinfo {author}
  {\bibfnamefont {R.-Q.}\ \bibnamefont {Wang}}, \bibinfo {author}
  {\bibfnamefont {P.}~\bibnamefont {Ye}}, \bibinfo {author} {\bibfnamefont
  {Y.}~\bibnamefont {Zhou}}, \bibinfo {author} {\bibfnamefont {G.-J.}\
  \bibnamefont {Fan-Yuan}}, \bibinfo {author} {\bibfnamefont {F.-X.}\
  \bibnamefont {Wang}}, \bibinfo {author} {\bibfnamefont {Y.-G.}\ \bibnamefont
  {Zhu}}, \emph {et~al.},\ }\bibfield  {title} {\bibinfo {title} {Twin-field
  quantum key distribution over 830-km fibre},\ }\href
  {https://doi.org/10.1038/s41566-021-00928-2} {\bibfield  {journal} {\bibinfo
  {journal} {Nat. Photon.}\ }\textbf {\bibinfo {volume} {16}},\ \bibinfo
  {pages} {154} (\bibinfo {year} {2022})}\BibitemShut {NoStop}%
\bibitem [{\citenamefont {Scheidl}\ \emph {et~al.}(2009)\citenamefont
  {Scheidl}, \citenamefont {Ursin}, \citenamefont {Fedrizzi}, \citenamefont
  {Ramelow}, \citenamefont {Ma}, \citenamefont {Herbst}, \citenamefont
  {Prevedel}, \citenamefont {Ratschbacher}, \citenamefont {Kofler},
  \citenamefont {Jennewein},\ and\ \citenamefont {Zeilinger}}]{Scheidl_2009}%
  \BibitemOpen
  \bibfield  {author} {\bibinfo {author} {\bibfnamefont {T.}~\bibnamefont
  {Scheidl}}, \bibinfo {author} {\bibfnamefont {R.}~\bibnamefont {Ursin}},
  \bibinfo {author} {\bibfnamefont {A.}~\bibnamefont {Fedrizzi}}, \bibinfo
  {author} {\bibfnamefont {S.}~\bibnamefont {Ramelow}}, \bibinfo {author}
  {\bibfnamefont {X.-S.}\ \bibnamefont {Ma}}, \bibinfo {author} {\bibfnamefont
  {T.}~\bibnamefont {Herbst}}, \bibinfo {author} {\bibfnamefont
  {R.}~\bibnamefont {Prevedel}}, \bibinfo {author} {\bibfnamefont
  {L.}~\bibnamefont {Ratschbacher}}, \bibinfo {author} {\bibfnamefont
  {J.}~\bibnamefont {Kofler}}, \bibinfo {author} {\bibfnamefont
  {T.}~\bibnamefont {Jennewein}},\ and\ \bibinfo {author} {\bibfnamefont
  {A.}~\bibnamefont {Zeilinger}},\ }\bibfield  {title} {\bibinfo {title}
  {Feasibility of 300 km quantum key distribution with entangled states},\
  }\href {https://doi.org/10.1088/1367-2630/11/8/085002} {\bibfield  {journal}
  {\bibinfo  {journal} {New J. Phys.}\ }\textbf {\bibinfo {volume} {11}},\
  \bibinfo {pages} {085002} (\bibinfo {year} {2009})}\BibitemShut {NoStop}%
\bibitem [{\citenamefont {Kržič}\ \emph {et~al.}(2023)\citenamefont
  {Kržič}, \citenamefont {Sharma}, \citenamefont {Spiess}, \citenamefont
  {Chandrashekara}, \citenamefont {Töpfer}, \citenamefont {Sauer},
  \citenamefont {del Campo}, \citenamefont {Kopf}, \citenamefont {Petscharnig},
  \citenamefont {Grafenauer}, \citenamefont {Lieger}, \citenamefont {Ömer},
  \citenamefont {Pacher}, \citenamefont {Berlich}, \citenamefont {Peschel},
  \citenamefont {Damm}, \citenamefont {Risse}, \citenamefont {Goy},
  \citenamefont {Rieländer}, \citenamefont {Tünnermann},\ and\ \citenamefont
  {Steinlechner}}]{krvzivc2022metropolitan}%
  \BibitemOpen
  \bibfield  {author} {\bibinfo {author} {\bibfnamefont {A.}~\bibnamefont
  {Kržič}}, \bibinfo {author} {\bibfnamefont {S.}~\bibnamefont {Sharma}},
  \bibinfo {author} {\bibfnamefont {C.}~\bibnamefont {Spiess}}, \bibinfo
  {author} {\bibfnamefont {U.}~\bibnamefont {Chandrashekara}}, \bibinfo
  {author} {\bibfnamefont {S.}~\bibnamefont {Töpfer}}, \bibinfo {author}
  {\bibfnamefont {G.}~\bibnamefont {Sauer}}, \bibinfo {author} {\bibfnamefont
  {L.~J. G.-M.}\ \bibnamefont {del Campo}}, \bibinfo {author} {\bibfnamefont
  {T.}~\bibnamefont {Kopf}}, \bibinfo {author} {\bibfnamefont {S.}~\bibnamefont
  {Petscharnig}}, \bibinfo {author} {\bibfnamefont {T.}~\bibnamefont
  {Grafenauer}}, \bibinfo {author} {\bibfnamefont {R.}~\bibnamefont {Lieger}},
  \bibinfo {author} {\bibfnamefont {B.}~\bibnamefont {Ömer}}, \bibinfo
  {author} {\bibfnamefont {C.}~\bibnamefont {Pacher}}, \bibinfo {author}
  {\bibfnamefont {R.}~\bibnamefont {Berlich}}, \bibinfo {author} {\bibfnamefont
  {T.}~\bibnamefont {Peschel}}, \bibinfo {author} {\bibfnamefont
  {C.}~\bibnamefont {Damm}}, \bibinfo {author} {\bibfnamefont {S.}~\bibnamefont
  {Risse}}, \bibinfo {author} {\bibfnamefont {M.}~\bibnamefont {Goy}}, \bibinfo
  {author} {\bibfnamefont {D.}~\bibnamefont {Rieländer}}, \bibinfo {author}
  {\bibfnamefont {A.}~\bibnamefont {Tünnermann}},\ and\ \bibinfo {author}
  {\bibfnamefont {F.}~\bibnamefont {Steinlechner}},\ }\bibfield  {title}
  {\bibinfo {title} {Towards metropolitan free-space quantum networks},\ }\href
  {https://www.nature.com/articles/s41534-023-00754-0#citeas} {\bibfield
  {journal} {\bibinfo  {journal} {npj Quantum Inf.}\ }\textbf {\bibinfo
  {volume} {9}},\ \bibinfo {pages} {95} (\bibinfo {year} {2023})}\BibitemShut
  {NoStop}%
\bibitem [{\citenamefont {Liao}\ \emph {et~al.}(2017)\citenamefont {Liao},
  \citenamefont {Cai}, \citenamefont {Liu}, \citenamefont {Zhang},
  \citenamefont {Li}, \citenamefont {Ren}, \citenamefont {Yin}, \citenamefont
  {Shen}, \citenamefont {Cao}, \citenamefont {Li} \emph
  {et~al.}}]{liao2017satellite}%
  \BibitemOpen
  \bibfield  {author} {\bibinfo {author} {\bibfnamefont {S.-K.}\ \bibnamefont
  {Liao}}, \bibinfo {author} {\bibfnamefont {W.-Q.}\ \bibnamefont {Cai}},
  \bibinfo {author} {\bibfnamefont {W.-Y.}\ \bibnamefont {Liu}}, \bibinfo
  {author} {\bibfnamefont {L.}~\bibnamefont {Zhang}}, \bibinfo {author}
  {\bibfnamefont {Y.}~\bibnamefont {Li}}, \bibinfo {author} {\bibfnamefont
  {J.-G.}\ \bibnamefont {Ren}}, \bibinfo {author} {\bibfnamefont
  {J.}~\bibnamefont {Yin}}, \bibinfo {author} {\bibfnamefont {Q.}~\bibnamefont
  {Shen}}, \bibinfo {author} {\bibfnamefont {Y.}~\bibnamefont {Cao}}, \bibinfo
  {author} {\bibfnamefont {Z.-P.}\ \bibnamefont {Li}}, \emph {et~al.},\
  }\bibfield  {title} {\bibinfo {title} {Satellite-to-ground quantum key
  distribution},\ }\href {https://doi.org/10.1038/nature23655} {\bibfield
  {journal} {\bibinfo  {journal} {Nature}\ }\textbf {\bibinfo {volume} {549}},\
  \bibinfo {pages} {43} (\bibinfo {year} {2017})}\BibitemShut {NoStop}%
\bibitem [{\citenamefont {Yin}\ \emph {et~al.}(2020)\citenamefont {Yin},
  \citenamefont {Li}, \citenamefont {Liao}, \citenamefont {Yang}, \citenamefont
  {Cao}, \citenamefont {Zhang}, \citenamefont {Ren}, \citenamefont {Cai},
  \citenamefont {Liu}, \citenamefont {Li} \emph
  {et~al.}}]{yin2020entanglement}%
  \BibitemOpen
  \bibfield  {author} {\bibinfo {author} {\bibfnamefont {J.}~\bibnamefont
  {Yin}}, \bibinfo {author} {\bibfnamefont {Y.-H.}\ \bibnamefont {Li}},
  \bibinfo {author} {\bibfnamefont {S.-K.}\ \bibnamefont {Liao}}, \bibinfo
  {author} {\bibfnamefont {M.}~\bibnamefont {Yang}}, \bibinfo {author}
  {\bibfnamefont {Y.}~\bibnamefont {Cao}}, \bibinfo {author} {\bibfnamefont
  {L.}~\bibnamefont {Zhang}}, \bibinfo {author} {\bibfnamefont {J.-G.}\
  \bibnamefont {Ren}}, \bibinfo {author} {\bibfnamefont {W.-Q.}\ \bibnamefont
  {Cai}}, \bibinfo {author} {\bibfnamefont {W.-Y.}\ \bibnamefont {Liu}},
  \bibinfo {author} {\bibfnamefont {S.-L.}\ \bibnamefont {Li}}, \emph
  {et~al.},\ }\bibfield  {title} {\bibinfo {title} {Entanglement-based secure
  quantum cryptography over 1,120 kilometres},\ }\href
  {https://doi.org/10.1038/s41586-020-2401-y} {\bibfield  {journal} {\bibinfo
  {journal} {Nature}\ }\textbf {\bibinfo {volume} {582}},\ \bibinfo {pages}
  {501} (\bibinfo {year} {2020})}\BibitemShut {NoStop}%
\bibitem [{\citenamefont {Boaron}\ \emph {et~al.}(2018)\citenamefont {Boaron},
  \citenamefont {Boso}, \citenamefont {Rusca}, \citenamefont {Vulliez},
  \citenamefont {Autebert}, \citenamefont {Caloz}, \citenamefont {Perrenoud},
  \citenamefont {Gras}, \citenamefont {Bussi\`eres}, \citenamefont {Li},
  \citenamefont {Nolan}, \citenamefont {Martin},\ and\ \citenamefont
  {Zbinden}}]{boaron2018quantum}%
  \BibitemOpen
  \bibfield  {author} {\bibinfo {author} {\bibfnamefont {A.}~\bibnamefont
  {Boaron}}, \bibinfo {author} {\bibfnamefont {G.}~\bibnamefont {Boso}},
  \bibinfo {author} {\bibfnamefont {D.}~\bibnamefont {Rusca}}, \bibinfo
  {author} {\bibfnamefont {C.}~\bibnamefont {Vulliez}}, \bibinfo {author}
  {\bibfnamefont {C.}~\bibnamefont {Autebert}}, \bibinfo {author}
  {\bibfnamefont {M.}~\bibnamefont {Caloz}}, \bibinfo {author} {\bibfnamefont
  {M.}~\bibnamefont {Perrenoud}}, \bibinfo {author} {\bibfnamefont
  {G.}~\bibnamefont {Gras}}, \bibinfo {author} {\bibfnamefont {F.}~\bibnamefont
  {Bussi\`eres}}, \bibinfo {author} {\bibfnamefont {M.-J.}\ \bibnamefont {Li}},
  \bibinfo {author} {\bibfnamefont {D.}~\bibnamefont {Nolan}}, \bibinfo
  {author} {\bibfnamefont {A.}~\bibnamefont {Martin}},\ and\ \bibinfo {author}
  {\bibfnamefont {H.}~\bibnamefont {Zbinden}},\ }\bibfield  {title} {\bibinfo
  {title} {Secure quantum key distribution over 421 km of optical fiber},\
  }\href {https://doi.org/10.1103/PhysRevLett.121.190502} {\bibfield  {journal}
  {\bibinfo  {journal} {Phys. Rev. Lett.}\ }\textbf {\bibinfo {volume} {121}},\
  \bibinfo {pages} {190502} (\bibinfo {year} {2018})}\BibitemShut {NoStop}%
\bibitem [{\citenamefont {Shi}\ \emph {et~al.}(2020)\citenamefont {Shi},
  \citenamefont {Thar}, \citenamefont {Poh}, \citenamefont {Grieve},
  \citenamefont {Kurtsiefer},\ and\ \citenamefont {Ling}}]{Shi2020}%
  \BibitemOpen
  \bibfield  {author} {\bibinfo {author} {\bibfnamefont {Y.}~\bibnamefont
  {Shi}}, \bibinfo {author} {\bibfnamefont {S.~M.}\ \bibnamefont {Thar}},
  \bibinfo {author} {\bibfnamefont {H.~S.}\ \bibnamefont {Poh}}, \bibinfo
  {author} {\bibfnamefont {J.~A.}\ \bibnamefont {Grieve}}, \bibinfo {author}
  {\bibfnamefont {C.}~\bibnamefont {Kurtsiefer}},\ and\ \bibinfo {author}
  {\bibfnamefont {A.}~\bibnamefont {Ling}},\ }\bibfield  {title} {\bibinfo
  {title} {Stable polarization entanglement based quantum key distribution over
  a deployed metropolitan fiber},\ }\href {https://doi.org/10.1063/5.0021755}
  {\bibfield  {journal} {\bibinfo  {journal} {Appl. Phys. Lett.}\ }\textbf
  {\bibinfo {volume} {117}},\ \bibinfo {pages} {124002} (\bibinfo {year}
  {2020})}\BibitemShut {NoStop}%
\bibitem [{\citenamefont {Chen}\ \emph {et~al.}(2022)\citenamefont {Chen},
  \citenamefont {Zhang}, \citenamefont {Liu}, \citenamefont {Jiang},
  \citenamefont {Zhao}, \citenamefont {Zhang}, \citenamefont {Chen},
  \citenamefont {Li}, \citenamefont {You}, \citenamefont {Wang}, \citenamefont
  {Chen}, \citenamefont {Wang}, \citenamefont {Zhang},\ and\ \citenamefont
  {Pan}}]{chen2022quantum}%
  \BibitemOpen
  \bibfield  {author} {\bibinfo {author} {\bibfnamefont {J.-P.}\ \bibnamefont
  {Chen}}, \bibinfo {author} {\bibfnamefont {C.}~\bibnamefont {Zhang}},
  \bibinfo {author} {\bibfnamefont {Y.}~\bibnamefont {Liu}}, \bibinfo {author}
  {\bibfnamefont {C.}~\bibnamefont {Jiang}}, \bibinfo {author} {\bibfnamefont
  {D.-F.}\ \bibnamefont {Zhao}}, \bibinfo {author} {\bibfnamefont {W.-J.}\
  \bibnamefont {Zhang}}, \bibinfo {author} {\bibfnamefont {F.-X.}\ \bibnamefont
  {Chen}}, \bibinfo {author} {\bibfnamefont {H.}~\bibnamefont {Li}}, \bibinfo
  {author} {\bibfnamefont {L.-X.}\ \bibnamefont {You}}, \bibinfo {author}
  {\bibfnamefont {Z.}~\bibnamefont {Wang}}, \bibinfo {author} {\bibfnamefont
  {Y.}~\bibnamefont {Chen}}, \bibinfo {author} {\bibfnamefont {X.-B.}\
  \bibnamefont {Wang}}, \bibinfo {author} {\bibfnamefont {Q.}~\bibnamefont
  {Zhang}},\ and\ \bibinfo {author} {\bibfnamefont {J.-W.}\ \bibnamefont
  {Pan}},\ }\bibfield  {title} {\bibinfo {title} {Quantum key distribution over
  658 km fiber with distributed vibration sensing},\ }\href
  {https://doi.org/10.1103/PhysRevLett.128.180502} {\bibfield  {journal}
  {\bibinfo  {journal} {Phys. Rev. Lett.}\ }\textbf {\bibinfo {volume} {128}},\
  \bibinfo {pages} {180502} (\bibinfo {year} {2022})}\BibitemShut {NoStop}%
\bibitem [{\citenamefont {Winzer}\ \emph {et~al.}(2018)\citenamefont {Winzer},
  \citenamefont {Neilson},\ and\ \citenamefont {Chraplyvy}}]{Winzer:18}%
  \BibitemOpen
  \bibfield  {author} {\bibinfo {author} {\bibfnamefont {P.~J.}\ \bibnamefont
  {Winzer}}, \bibinfo {author} {\bibfnamefont {D.~T.}\ \bibnamefont
  {Neilson}},\ and\ \bibinfo {author} {\bibfnamefont {A.~R.}\ \bibnamefont
  {Chraplyvy}},\ }\bibfield  {title} {\bibinfo {title} {Fiber-optic
  transmission and networking: the previous 20 and the next 20 years
  \[invited\]},\ }\href {https://doi.org/10.1364/OE.26.024190} {\bibfield
  {journal} {\bibinfo  {journal} {Opt. Express}\ }\textbf {\bibinfo {volume}
  {26}},\ \bibinfo {pages} {24190} (\bibinfo {year} {2018})}\BibitemShut
  {NoStop}%
\bibitem [{\citenamefont {Ribezzo}\ \emph {et~al.}(2023)\citenamefont
  {Ribezzo}, \citenamefont {Zahidy}, \citenamefont {Vagniluca}, \citenamefont
  {Biagi}, \citenamefont {Francesconi}, \citenamefont {Occhipinti},
  \citenamefont {Oxenl{\o}we}, \citenamefont {Lon{\v{c}}ari{\'c}},
  \citenamefont {Cviti{\'c}}, \citenamefont {Stip{\v{c}}evi{\'c}},
  \citenamefont {Pu{\v{s}}avec}, \citenamefont {Kaltenbaek}, \citenamefont
  {Ram{\v{s}}ak}, \citenamefont {Cesa}, \citenamefont {Giorgetti},
  \citenamefont {Scazza}, \citenamefont {Bassi}, \citenamefont {De~Natale},
  \citenamefont {Cataliotti}, \citenamefont {Inguscio}, \citenamefont {Bacco},\
  and\ \citenamefont {Zavatta}}]{ribezzo2022deploying}%
  \BibitemOpen
  \bibfield  {author} {\bibinfo {author} {\bibfnamefont {D.}~\bibnamefont
  {Ribezzo}}, \bibinfo {author} {\bibfnamefont {M.}~\bibnamefont {Zahidy}},
  \bibinfo {author} {\bibfnamefont {I.}~\bibnamefont {Vagniluca}}, \bibinfo
  {author} {\bibfnamefont {N.}~\bibnamefont {Biagi}}, \bibinfo {author}
  {\bibfnamefont {S.}~\bibnamefont {Francesconi}}, \bibinfo {author}
  {\bibfnamefont {T.}~\bibnamefont {Occhipinti}}, \bibinfo {author}
  {\bibfnamefont {L.~K.}\ \bibnamefont {Oxenl{\o}we}}, \bibinfo {author}
  {\bibfnamefont {M.}~\bibnamefont {Lon{\v{c}}ari{\'c}}}, \bibinfo {author}
  {\bibfnamefont {I.}~\bibnamefont {Cviti{\'c}}}, \bibinfo {author}
  {\bibfnamefont {M.}~\bibnamefont {Stip{\v{c}}evi{\'c}}}, \bibinfo {author}
  {\bibfnamefont {{\v{Z}}.}~\bibnamefont {Pu{\v{s}}avec}}, \bibinfo {author}
  {\bibfnamefont {R.}~\bibnamefont {Kaltenbaek}}, \bibinfo {author}
  {\bibfnamefont {A.}~\bibnamefont {Ram{\v{s}}ak}}, \bibinfo {author}
  {\bibfnamefont {F.}~\bibnamefont {Cesa}}, \bibinfo {author} {\bibfnamefont
  {G.}~\bibnamefont {Giorgetti}}, \bibinfo {author} {\bibfnamefont
  {F.}~\bibnamefont {Scazza}}, \bibinfo {author} {\bibfnamefont
  {A.}~\bibnamefont {Bassi}}, \bibinfo {author} {\bibfnamefont
  {P.}~\bibnamefont {De~Natale}}, \bibinfo {author} {\bibfnamefont {F.~S.}\
  \bibnamefont {Cataliotti}}, \bibinfo {author} {\bibfnamefont
  {M.}~\bibnamefont {Inguscio}}, \bibinfo {author} {\bibfnamefont
  {D.}~\bibnamefont {Bacco}},\ and\ \bibinfo {author} {\bibfnamefont
  {A.}~\bibnamefont {Zavatta}},\ }\bibfield  {title} {\bibinfo {title}
  {Deploying an inter-european quantum network},\ }\href
  {https://onlinelibrary.wiley.com/doi/full/10.1002/qute.202200061} {\bibfield
  {journal} {\bibinfo  {journal} {Adv. Quantum Technol.}\ }\textbf {\bibinfo
  {volume} {6}},\ \bibinfo {pages} {2200061} (\bibinfo {year}
  {2023})}\BibitemShut {NoStop}%
\bibitem [{\citenamefont {Wang}\ \emph {et~al.}(2020)\citenamefont {Wang},
  \citenamefont {Mao}, \citenamefont {Shen}, \citenamefont {Zhang},
  \citenamefont {Lan}, \citenamefont {Ge}, \citenamefont {Gao}, \citenamefont
  {Li}, \citenamefont {Tang}, \citenamefont {Tang}, \citenamefont {Zhang},
  \citenamefont {Chen},\ and\ \citenamefont {Pan}}]{Wang:20}%
  \BibitemOpen
  \bibfield  {author} {\bibinfo {author} {\bibfnamefont {B.-X.}\ \bibnamefont
  {Wang}}, \bibinfo {author} {\bibfnamefont {Y.}~\bibnamefont {Mao}}, \bibinfo
  {author} {\bibfnamefont {L.}~\bibnamefont {Shen}}, \bibinfo {author}
  {\bibfnamefont {L.}~\bibnamefont {Zhang}}, \bibinfo {author} {\bibfnamefont
  {X.-B.}\ \bibnamefont {Lan}}, \bibinfo {author} {\bibfnamefont
  {D.}~\bibnamefont {Ge}}, \bibinfo {author} {\bibfnamefont {Y.}~\bibnamefont
  {Gao}}, \bibinfo {author} {\bibfnamefont {J.}~\bibnamefont {Li}}, \bibinfo
  {author} {\bibfnamefont {Y.-L.}\ \bibnamefont {Tang}}, \bibinfo {author}
  {\bibfnamefont {S.-B.}\ \bibnamefont {Tang}}, \bibinfo {author}
  {\bibfnamefont {J.}~\bibnamefont {Zhang}}, \bibinfo {author} {\bibfnamefont
  {T.-Y.}\ \bibnamefont {Chen}},\ and\ \bibinfo {author} {\bibfnamefont
  {J.-W.}\ \bibnamefont {Pan}},\ }\bibfield  {title} {\bibinfo {title}
  {Long-distance transmission of quantum key distribution coexisting with
  classical optical communication over a weakly-coupled few-mode fiber},\
  }\href {https://doi.org/10.1364/OE.388857} {\bibfield  {journal} {\bibinfo
  {journal} {Opt. Express}\ }\textbf {\bibinfo {volume} {28}},\ \bibinfo
  {pages} {12558} (\bibinfo {year} {2020})}\BibitemShut {NoStop}%
\bibitem [{\citenamefont {Alarc\'on}\ \emph {et~al.}(2021)\citenamefont
  {Alarc\'on}, \citenamefont {Argillander}, \citenamefont {Lima},\ and\
  \citenamefont {Xavier}}]{alarcon2021few}%
  \BibitemOpen
  \bibfield  {author} {\bibinfo {author} {\bibfnamefont {A.}~\bibnamefont
  {Alarc\'on}}, \bibinfo {author} {\bibfnamefont {J.}~\bibnamefont
  {Argillander}}, \bibinfo {author} {\bibfnamefont {G.}~\bibnamefont {Lima}},\
  and\ \bibinfo {author} {\bibfnamefont {G.}~\bibnamefont {Xavier}},\
  }\bibfield  {title} {\bibinfo {title} {Few-mode-fiber technology fine-tunes
  losses in quantum communication systems},\ }\href
  {https://doi.org/10.1103/PhysRevApplied.16.034018} {\bibfield  {journal}
  {\bibinfo  {journal} {Phys. Rev. Applied}\ }\textbf {\bibinfo {volume}
  {16}},\ \bibinfo {pages} {034018} (\bibinfo {year} {2021})}\BibitemShut
  {NoStop}%
\bibitem [{\citenamefont {Zhou}\ \emph {et~al.}(2021)\citenamefont {Zhou},
  \citenamefont {Braverman}, \citenamefont {Fyffe}, \citenamefont {Zhang},
  \citenamefont {Zhao}, \citenamefont {Willner}, \citenamefont {Shi},\ and\
  \citenamefont {Boyd}}]{zhou2021high}%
  \BibitemOpen
  \bibfield  {author} {\bibinfo {author} {\bibfnamefont {Y.}~\bibnamefont
  {Zhou}}, \bibinfo {author} {\bibfnamefont {B.}~\bibnamefont {Braverman}},
  \bibinfo {author} {\bibfnamefont {A.}~\bibnamefont {Fyffe}}, \bibinfo
  {author} {\bibfnamefont {R.}~\bibnamefont {Zhang}}, \bibinfo {author}
  {\bibfnamefont {J.}~\bibnamefont {Zhao}}, \bibinfo {author} {\bibfnamefont
  {A.~E.}\ \bibnamefont {Willner}}, \bibinfo {author} {\bibfnamefont
  {Z.}~\bibnamefont {Shi}},\ and\ \bibinfo {author} {\bibfnamefont {R.~W.}\
  \bibnamefont {Boyd}},\ }\bibfield  {title} {\bibinfo {title} {High-fidelity
  spatial mode transmission through a 1-km-long multimode fiber via vectorial
  time reversal},\ }\href {https://doi.org/10.1038/s41467-021-22071-w}
  {\bibfield  {journal} {\bibinfo  {journal} {Nat. Commun.}\ }\textbf {\bibinfo
  {volume} {12}},\ \bibinfo {pages} {1866} (\bibinfo {year}
  {2021})}\BibitemShut {NoStop}%
\bibitem [{\citenamefont {Amitonova}\ \emph {et~al.}(2020)\citenamefont
  {Amitonova}, \citenamefont {Tentrup}, \citenamefont {Vellekoop},\ and\
  \citenamefont {Pinkse}}]{amitonova2020quantum}%
  \BibitemOpen
  \bibfield  {author} {\bibinfo {author} {\bibfnamefont {L.~V.}\ \bibnamefont
  {Amitonova}}, \bibinfo {author} {\bibfnamefont {T.~B.}\ \bibnamefont
  {Tentrup}}, \bibinfo {author} {\bibfnamefont {I.~M.}\ \bibnamefont
  {Vellekoop}},\ and\ \bibinfo {author} {\bibfnamefont {P.~W.}\ \bibnamefont
  {Pinkse}},\ }\bibfield  {title} {\bibinfo {title} {Quantum key establishment
  via a multimode fiber},\ }\href {https://doi.org/10.1364/OE.380791}
  {\bibfield  {journal} {\bibinfo  {journal} {Opt. Express}\ }\textbf {\bibinfo
  {volume} {28}},\ \bibinfo {pages} {5965} (\bibinfo {year}
  {2020})}\BibitemShut {NoStop}%
\bibitem [{\citenamefont {Sit}\ \emph {et~al.}(2018)\citenamefont {Sit},
  \citenamefont {Fickler}, \citenamefont {Alsaiari}, \citenamefont {Bouchard},
  \citenamefont {Larocque}, \citenamefont {Gregg}, \citenamefont {Yan},
  \citenamefont {Boyd}, \citenamefont {Ramachandran},\ and\ \citenamefont
  {Karimi}}]{sit2018quantum}%
  \BibitemOpen
  \bibfield  {author} {\bibinfo {author} {\bibfnamefont {A.}~\bibnamefont
  {Sit}}, \bibinfo {author} {\bibfnamefont {R.}~\bibnamefont {Fickler}},
  \bibinfo {author} {\bibfnamefont {F.}~\bibnamefont {Alsaiari}}, \bibinfo
  {author} {\bibfnamefont {F.}~\bibnamefont {Bouchard}}, \bibinfo {author}
  {\bibfnamefont {H.}~\bibnamefont {Larocque}}, \bibinfo {author}
  {\bibfnamefont {P.}~\bibnamefont {Gregg}}, \bibinfo {author} {\bibfnamefont
  {L.}~\bibnamefont {Yan}}, \bibinfo {author} {\bibfnamefont {R.~W.}\
  \bibnamefont {Boyd}}, \bibinfo {author} {\bibfnamefont {S.}~\bibnamefont
  {Ramachandran}},\ and\ \bibinfo {author} {\bibfnamefont {E.}~\bibnamefont
  {Karimi}},\ }\bibfield  {title} {\bibinfo {title} {Quantum cryptography with
  structured photons through a vortex fiber},\ }\href
  {https://doi.org/10.1364/OL.43.004108} {\bibfield  {journal} {\bibinfo
  {journal} {Opt. Lett.}\ }\textbf {\bibinfo {volume} {43}},\ \bibinfo {pages}
  {4108} (\bibinfo {year} {2018})}\BibitemShut {NoStop}%
\bibitem [{\citenamefont {Dynes}\ \emph {et~al.}(2016)\citenamefont {Dynes},
  \citenamefont {Kindness}, \citenamefont {Tam}, \citenamefont {Plews},
  \citenamefont {Sharpe}, \citenamefont {Lucamarini}, \citenamefont
  {Fr\"{o}hlich}, \citenamefont {Yuan}, \citenamefont {Penty},\ and\
  \citenamefont {Shields}}]{Dynes:16}%
  \BibitemOpen
  \bibfield  {author} {\bibinfo {author} {\bibfnamefont {J.~F.}\ \bibnamefont
  {Dynes}}, \bibinfo {author} {\bibfnamefont {S.~J.}\ \bibnamefont {Kindness}},
  \bibinfo {author} {\bibfnamefont {S.~W.-B.}\ \bibnamefont {Tam}}, \bibinfo
  {author} {\bibfnamefont {A.}~\bibnamefont {Plews}}, \bibinfo {author}
  {\bibfnamefont {A.~W.}\ \bibnamefont {Sharpe}}, \bibinfo {author}
  {\bibfnamefont {M.}~\bibnamefont {Lucamarini}}, \bibinfo {author}
  {\bibfnamefont {B.}~\bibnamefont {Fr\"{o}hlich}}, \bibinfo {author}
  {\bibfnamefont {Z.~L.}\ \bibnamefont {Yuan}}, \bibinfo {author}
  {\bibfnamefont {R.~V.}\ \bibnamefont {Penty}},\ and\ \bibinfo {author}
  {\bibfnamefont {A.~J.}\ \bibnamefont {Shields}},\ }\bibfield  {title}
  {\bibinfo {title} {Quantum key distribution over multicore fiber},\ }\href
  {https://doi.org/10.1364/OE.24.008081} {\bibfield  {journal} {\bibinfo
  {journal} {Opt. Express}\ }\textbf {\bibinfo {volume} {24}},\ \bibinfo
  {pages} {8081} (\bibinfo {year} {2016})}\BibitemShut {NoStop}%
\bibitem [{\citenamefont {Bacco}\ \emph {et~al.}(2017)\citenamefont {Bacco},
  \citenamefont {Ding}, \citenamefont {Dalgaard}, \citenamefont {Rottwitt},\
  and\ \citenamefont {Oxenl{\o}we}}]{bacco2017space}%
  \BibitemOpen
  \bibfield  {author} {\bibinfo {author} {\bibfnamefont {D.}~\bibnamefont
  {Bacco}}, \bibinfo {author} {\bibfnamefont {Y.}~\bibnamefont {Ding}},
  \bibinfo {author} {\bibfnamefont {K.}~\bibnamefont {Dalgaard}}, \bibinfo
  {author} {\bibfnamefont {K.}~\bibnamefont {Rottwitt}},\ and\ \bibinfo
  {author} {\bibfnamefont {L.~K.}\ \bibnamefont {Oxenl{\o}we}},\ }\bibfield
  {title} {\bibinfo {title} {Space division multiplexing chip-to-chip quantum
  key distribution},\ }\href {https://doi.org/10.1038/s41598-017-12309-3}
  {\bibfield  {journal} {\bibinfo  {journal} {Sci. Rep.}\ }\textbf {\bibinfo
  {volume} {7}},\ \bibinfo {pages} {1} (\bibinfo {year} {2017})}\BibitemShut
  {NoStop}%
\bibitem [{\citenamefont {Wengerowsky}\ \emph {et~al.}(2020)\citenamefont
  {Wengerowsky}, \citenamefont {Joshi}, \citenamefont {Steinlechner},
  \citenamefont {Zichi}, \citenamefont {Liu}, \citenamefont {Scheidl},
  \citenamefont {Dobrovolskiy}, \citenamefont {van~der Molen}, \citenamefont
  {Los}, \citenamefont {Zwiller}, \citenamefont {Versteegh}, \citenamefont
  {Mura}, \citenamefont {Calonico}, \citenamefont {Inguscio}, \citenamefont
  {Zeilinger}, \citenamefont {Xuereb},\ and\ \citenamefont
  {Ursin}}]{Wengerowsky2020}%
  \BibitemOpen
  \bibfield  {author} {\bibinfo {author} {\bibfnamefont {S.}~\bibnamefont
  {Wengerowsky}}, \bibinfo {author} {\bibfnamefont {S.~K.}\ \bibnamefont
  {Joshi}}, \bibinfo {author} {\bibfnamefont {F.}~\bibnamefont {Steinlechner}},
  \bibinfo {author} {\bibfnamefont {J.~R.}\ \bibnamefont {Zichi}}, \bibinfo
  {author} {\bibfnamefont {B.}~\bibnamefont {Liu}}, \bibinfo {author}
  {\bibfnamefont {T.}~\bibnamefont {Scheidl}}, \bibinfo {author} {\bibfnamefont
  {S.~M.}\ \bibnamefont {Dobrovolskiy}}, \bibinfo {author} {\bibfnamefont
  {R.}~\bibnamefont {van~der Molen}}, \bibinfo {author} {\bibfnamefont
  {J.~W.~N.}\ \bibnamefont {Los}}, \bibinfo {author} {\bibfnamefont
  {V.}~\bibnamefont {Zwiller}}, \bibinfo {author} {\bibfnamefont {M.~A.~M.}\
  \bibnamefont {Versteegh}}, \bibinfo {author} {\bibfnamefont {A.}~\bibnamefont
  {Mura}}, \bibinfo {author} {\bibfnamefont {D.}~\bibnamefont {Calonico}},
  \bibinfo {author} {\bibfnamefont {M.}~\bibnamefont {Inguscio}}, \bibinfo
  {author} {\bibfnamefont {A.}~\bibnamefont {Zeilinger}}, \bibinfo {author}
  {\bibfnamefont {A.}~\bibnamefont {Xuereb}},\ and\ \bibinfo {author}
  {\bibfnamefont {R.}~\bibnamefont {Ursin}},\ }\bibfield  {title} {\bibinfo
  {title} {Passively stable distribution of polarisation entanglement over 192
  km of deployed optical fibre},\ }\href
  {https://doi.org/10.1038/s41534-019-0238-8} {\bibfield  {journal} {\bibinfo
  {journal} {npj Quantum Inf.}\ }\textbf {\bibinfo {volume} {6}},\ \bibinfo
  {pages} {5} (\bibinfo {year} {2020})}\BibitemShut {NoStop}%
\bibitem [{\citenamefont {Neumann}\ \emph
  {et~al.}(2022{\natexlab{a}})\citenamefont {Neumann}, \citenamefont {Buchner},
  \citenamefont {Bulla}, \citenamefont {Bohmann},\ and\ \citenamefont
  {Ursin}}]{neumann2022quapital}%
  \BibitemOpen
  \bibfield  {author} {\bibinfo {author} {\bibfnamefont {S.~P.}\ \bibnamefont
  {Neumann}}, \bibinfo {author} {\bibfnamefont {A.}~\bibnamefont {Buchner}},
  \bibinfo {author} {\bibfnamefont {L.}~\bibnamefont {Bulla}}, \bibinfo
  {author} {\bibfnamefont {M.}~\bibnamefont {Bohmann}},\ and\ \bibinfo {author}
  {\bibfnamefont {R.}~\bibnamefont {Ursin}},\ }\bibfield  {title} {\bibinfo
  {title} {Continuous entanglement distribution over a transnational
  248{\hspace{0.167em}}km fiber link},\ }\href
  {https://doi.org/10.1038/s41467-022-33919-0} {\bibfield  {journal} {\bibinfo
  {journal} {Nat. Commun.}\ }\textbf {\bibinfo {volume} {13}},\ \bibinfo
  {pages} {6134} (\bibinfo {year} {2022}{\natexlab{a}})}\BibitemShut {NoStop}%
\bibitem [{\citenamefont {Neumann}\ \emph
  {et~al.}(2022{\natexlab{b}})\citenamefont {Neumann}, \citenamefont
  {Selimovic}, \citenamefont {Bohmann},\ and\ \citenamefont
  {Ursin}}]{neumann2022experimental}%
  \BibitemOpen
  \bibfield  {author} {\bibinfo {author} {\bibfnamefont {S.~P.}\ \bibnamefont
  {Neumann}}, \bibinfo {author} {\bibfnamefont {M.}~\bibnamefont {Selimovic}},
  \bibinfo {author} {\bibfnamefont {M.}~\bibnamefont {Bohmann}},\ and\ \bibinfo
  {author} {\bibfnamefont {R.}~\bibnamefont {Ursin}},\ }\bibfield  {title}
  {\bibinfo {title} {Experimental entanglement generation for quantum key
  distribution beyond 1 gbit/s},\ }\href
  {https://doi.org/10.22331/q-2022-09-29-822} {\bibfield  {journal} {\bibinfo
  {journal} {Quantum}\ }\textbf {\bibinfo {volume} {6}},\ \bibinfo {pages}
  {822} (\bibinfo {year} {2022}{\natexlab{b}})}\BibitemShut {NoStop}%
\bibitem [{\citenamefont {Brambila}\ \emph {et~al.}(2023)\citenamefont
  {Brambila}, \citenamefont {G{\'o}mez}, \citenamefont {Fazili}, \citenamefont
  {Gr{\"a}fe},\ and\ \citenamefont {Steinlechner}}]{brambila2023ultrabright}%
  \BibitemOpen
  \bibfield  {author} {\bibinfo {author} {\bibfnamefont {E.}~\bibnamefont
  {Brambila}}, \bibinfo {author} {\bibfnamefont {R.}~\bibnamefont {G{\'o}mez}},
  \bibinfo {author} {\bibfnamefont {R.}~\bibnamefont {Fazili}}, \bibinfo
  {author} {\bibfnamefont {M.}~\bibnamefont {Gr{\"a}fe}},\ and\ \bibinfo
  {author} {\bibfnamefont {F.}~\bibnamefont {Steinlechner}},\ }\bibfield
  {title} {\bibinfo {title} {Ultrabright polarization-entangled photon pair
  source for frequency-multiplexed quantum communication in free-space},\
  }\href {https://doi.org/10.1364/OE.461802} {\bibfield  {journal} {\bibinfo
  {journal} {Opt. Express}\ }\textbf {\bibinfo {volume} {31}},\ \bibinfo
  {pages} {16107} (\bibinfo {year} {2023})}\BibitemShut {NoStop}%
\bibitem [{\citenamefont {Wengerowsky}\ \emph {et~al.}(2018)\citenamefont
  {Wengerowsky}, \citenamefont {Joshi}, \citenamefont {Steinlechner},
  \citenamefont {H{\"u}bel},\ and\ \citenamefont
  {Ursin}}]{wengerowsky2018entanglement}%
  \BibitemOpen
  \bibfield  {author} {\bibinfo {author} {\bibfnamefont {S.}~\bibnamefont
  {Wengerowsky}}, \bibinfo {author} {\bibfnamefont {S.~K.}\ \bibnamefont
  {Joshi}}, \bibinfo {author} {\bibfnamefont {F.}~\bibnamefont {Steinlechner}},
  \bibinfo {author} {\bibfnamefont {H.}~\bibnamefont {H{\"u}bel}},\ and\
  \bibinfo {author} {\bibfnamefont {R.}~\bibnamefont {Ursin}},\ }\bibfield
  {title} {\bibinfo {title} {An entanglement-based wavelength-multiplexed
  quantum communication network},\ }\href
  {https://doi.org/https://doi.org/10.1038/s41586-018-0766-y} {\bibfield
  {journal} {\bibinfo  {journal} {Nature}\ }\textbf {\bibinfo {volume} {564}},\
  \bibinfo {pages} {225} (\bibinfo {year} {2018})}\BibitemShut {NoStop}%
\bibitem [{\citenamefont {Joshi}\ \emph {et~al.}(2020)\citenamefont {Joshi},
  \citenamefont {Aktas}, \citenamefont {Wengerowsky}, \citenamefont
  {Lon{\v{c}}ari{\'c}}, \citenamefont {Neumann}, \citenamefont {Liu},
  \citenamefont {Scheidl}, \citenamefont {Lorenzo}, \citenamefont {Samec},
  \citenamefont {Kling} \emph {et~al.}}]{joshi2020trusted}%
  \BibitemOpen
  \bibfield  {author} {\bibinfo {author} {\bibfnamefont {S.~K.}\ \bibnamefont
  {Joshi}}, \bibinfo {author} {\bibfnamefont {D.}~\bibnamefont {Aktas}},
  \bibinfo {author} {\bibfnamefont {S.}~\bibnamefont {Wengerowsky}}, \bibinfo
  {author} {\bibfnamefont {M.}~\bibnamefont {Lon{\v{c}}ari{\'c}}}, \bibinfo
  {author} {\bibfnamefont {S.~P.}\ \bibnamefont {Neumann}}, \bibinfo {author}
  {\bibfnamefont {B.}~\bibnamefont {Liu}}, \bibinfo {author} {\bibfnamefont
  {T.}~\bibnamefont {Scheidl}}, \bibinfo {author} {\bibfnamefont {G.~C.}\
  \bibnamefont {Lorenzo}}, \bibinfo {author} {\bibfnamefont
  {{\v{Z}}.}~\bibnamefont {Samec}}, \bibinfo {author} {\bibfnamefont
  {L.}~\bibnamefont {Kling}}, \emph {et~al.},\ }\bibfield  {title} {\bibinfo
  {title} {A trusted node--free eight-user metropolitan quantum communication
  network},\ }\href {https://doi.org/https://doi.org/10.1126/sciadv.aba0959}
  {\bibfield  {journal} {\bibinfo  {journal} {Sci. Adv.}\ }\textbf {\bibinfo
  {volume} {6}},\ \bibinfo {pages} {eaba0959} (\bibinfo {year}
  {2020})}\BibitemShut {NoStop}%
\bibitem [{\citenamefont {Richardson}\ \emph {et~al.}(2013)\citenamefont
  {Richardson}, \citenamefont {Fini},\ and\ \citenamefont
  {Nelson}}]{richardson2013}%
  \BibitemOpen
  \bibfield  {author} {\bibinfo {author} {\bibfnamefont {D.~J.}\ \bibnamefont
  {Richardson}}, \bibinfo {author} {\bibfnamefont {J.~M.}\ \bibnamefont
  {Fini}},\ and\ \bibinfo {author} {\bibfnamefont {L.~E.}\ \bibnamefont
  {Nelson}},\ }\bibfield  {title} {\bibinfo {title} {Space-division
  multiplexing in optical fibres},\ }\href
  {https://doi.org/10.1038/nphoton.2013.94} {\bibfield  {journal} {\bibinfo
  {journal} {Nature Photon.}\ }\textbf {\bibinfo {volume} {7}},\ \bibinfo
  {pages} {354} (\bibinfo {year} {2013})}\BibitemShut {NoStop}%
\bibitem [{\citenamefont {Puttnam}\ \emph {et~al.}(2021)\citenamefont
  {Puttnam}, \citenamefont {Rademacher},\ and\ \citenamefont
  {Lu\'{i}s}}]{Puttnam:21}%
  \BibitemOpen
  \bibfield  {author} {\bibinfo {author} {\bibfnamefont {B.~J.}\ \bibnamefont
  {Puttnam}}, \bibinfo {author} {\bibfnamefont {G.}~\bibnamefont
  {Rademacher}},\ and\ \bibinfo {author} {\bibfnamefont {R.~S.}\ \bibnamefont
  {Lu\'{i}s}},\ }\bibfield  {title} {\bibinfo {title} {Space-division
  multiplexing for optical fiber communications},\ }\href
  {https://doi.org/10.1364/OPTICA.427631} {\bibfield  {journal} {\bibinfo
  {journal} {Optica}\ }\textbf {\bibinfo {volume} {8}},\ \bibinfo {pages}
  {1186} (\bibinfo {year} {2021})}\BibitemShut {NoStop}%
\bibitem [{\citenamefont {Xavier}\ and\ \citenamefont
  {Lima}(2020)}]{xavier2020quantum}%
  \BibitemOpen
  \bibfield  {author} {\bibinfo {author} {\bibfnamefont {G.~B.}\ \bibnamefont
  {Xavier}}\ and\ \bibinfo {author} {\bibfnamefont {G.}~\bibnamefont {Lima}},\
  }\bibfield  {title} {\bibinfo {title} {Quantum information processing with
  space-division multiplexing optical fibres},\ }\href
  {https://doi.org/10.1038/s42005-019-0269-7} {\bibfield  {journal} {\bibinfo
  {journal} {Commun. Phys.}\ }\textbf {\bibinfo {volume} {3}},\ \bibinfo
  {pages} {9} (\bibinfo {year} {2020})}\BibitemShut {NoStop}%
\bibitem [{\citenamefont {Bacco}\ \emph {et~al.}(2019)\citenamefont {Bacco},
  \citenamefont {Lio}, \citenamefont {Cozzolino}, \citenamefont {Ros},
  \citenamefont {Guo}, \citenamefont {Ding}, \citenamefont {Sasaki},
  \citenamefont {Aikawa}, \citenamefont {Miki}, \citenamefont {Terai},
  \citenamefont {Yamashita}, \citenamefont {Neergaard-Nielsen}, \citenamefont
  {Galili}, \citenamefont {Rottwitt}, \citenamefont {Andersen}, \citenamefont
  {Morioka},\ and\ \citenamefont {Oxenl{\o}we}}]{Bacco2019}%
  \BibitemOpen
  \bibfield  {author} {\bibinfo {author} {\bibfnamefont {D.}~\bibnamefont
  {Bacco}}, \bibinfo {author} {\bibfnamefont {B.~D.}\ \bibnamefont {Lio}},
  \bibinfo {author} {\bibfnamefont {D.}~\bibnamefont {Cozzolino}}, \bibinfo
  {author} {\bibfnamefont {F.~D.}\ \bibnamefont {Ros}}, \bibinfo {author}
  {\bibfnamefont {X.}~\bibnamefont {Guo}}, \bibinfo {author} {\bibfnamefont
  {Y.}~\bibnamefont {Ding}}, \bibinfo {author} {\bibfnamefont {Y.}~\bibnamefont
  {Sasaki}}, \bibinfo {author} {\bibfnamefont {K.}~\bibnamefont {Aikawa}},
  \bibinfo {author} {\bibfnamefont {S.}~\bibnamefont {Miki}}, \bibinfo {author}
  {\bibfnamefont {H.}~\bibnamefont {Terai}}, \bibinfo {author} {\bibfnamefont
  {T.}~\bibnamefont {Yamashita}}, \bibinfo {author} {\bibfnamefont {J.~S.}\
  \bibnamefont {Neergaard-Nielsen}}, \bibinfo {author} {\bibfnamefont
  {M.}~\bibnamefont {Galili}}, \bibinfo {author} {\bibfnamefont
  {K.}~\bibnamefont {Rottwitt}}, \bibinfo {author} {\bibfnamefont {U.~L.}\
  \bibnamefont {Andersen}}, \bibinfo {author} {\bibfnamefont {T.}~\bibnamefont
  {Morioka}},\ and\ \bibinfo {author} {\bibfnamefont {L.~K.}\ \bibnamefont
  {Oxenl{\o}we}},\ }\bibfield  {title} {\bibinfo {title} {Boosting the secret
  key rate in a shared quantum and classical fibre communication system},\
  }\href {https://doi.org/10.1038/s42005-019-0238-1} {\bibfield  {journal}
  {\bibinfo  {journal} {Commun. Phys.}\ }\textbf {\bibinfo {volume} {2}},\
  \bibinfo {pages} {140} (\bibinfo {year} {2019})}\BibitemShut {NoStop}%
\bibitem [{\citenamefont {Ca\~nas}\ \emph {et~al.}(2017)\citenamefont
  {Ca\~nas}, \citenamefont {Vera}, \citenamefont {Cari\~ne}, \citenamefont
  {Gonz\'alez}, \citenamefont {Cardenas}, \citenamefont {Connolly},
  \citenamefont {Przysiezna}, \citenamefont {G\'omez}, \citenamefont
  {Figueroa}, \citenamefont {Vallone}, \citenamefont {Villoresi}, \citenamefont
  {da~Silva}, \citenamefont {Xavier},\ and\ \citenamefont
  {Lima}}]{canas2017decoy}%
  \BibitemOpen
  \bibfield  {author} {\bibinfo {author} {\bibfnamefont {G.}~\bibnamefont
  {Ca\~nas}}, \bibinfo {author} {\bibfnamefont {N.}~\bibnamefont {Vera}},
  \bibinfo {author} {\bibfnamefont {J.}~\bibnamefont {Cari\~ne}}, \bibinfo
  {author} {\bibfnamefont {P.}~\bibnamefont {Gonz\'alez}}, \bibinfo {author}
  {\bibfnamefont {J.}~\bibnamefont {Cardenas}}, \bibinfo {author}
  {\bibfnamefont {P.~W.~R.}\ \bibnamefont {Connolly}}, \bibinfo {author}
  {\bibfnamefont {A.}~\bibnamefont {Przysiezna}}, \bibinfo {author}
  {\bibfnamefont {E.~S.}\ \bibnamefont {G\'omez}}, \bibinfo {author}
  {\bibfnamefont {M.}~\bibnamefont {Figueroa}}, \bibinfo {author}
  {\bibfnamefont {G.}~\bibnamefont {Vallone}}, \bibinfo {author} {\bibfnamefont
  {P.}~\bibnamefont {Villoresi}}, \bibinfo {author} {\bibfnamefont {T.~F.}\
  \bibnamefont {da~Silva}}, \bibinfo {author} {\bibfnamefont {G.~B.}\
  \bibnamefont {Xavier}},\ and\ \bibinfo {author} {\bibfnamefont
  {G.}~\bibnamefont {Lima}},\ }\bibfield  {title} {\bibinfo {title}
  {High-dimensional decoy-state quantum key distribution over multicore
  telecommunication fibers},\ }\href
  {https://doi.org/10.1103/PhysRevA.96.022317} {\bibfield  {journal} {\bibinfo
  {journal} {Phys. Rev. A}\ }\textbf {\bibinfo {volume} {96}},\ \bibinfo
  {pages} {022317} (\bibinfo {year} {2017})}\BibitemShut {NoStop}%
\bibitem [{\citenamefont {Ortega}\ \emph {et~al.}(2021)\citenamefont {Ortega},
  \citenamefont {Dovzhik}, \citenamefont {Fuenzalida}, \citenamefont
  {Wengerowsky}, \citenamefont {Alvarado-Zacarias}, \citenamefont {Shiozaki},
  \citenamefont {Amezcua-Correa}, \citenamefont {Bohmann},\ and\ \citenamefont
  {Ursin}}]{ortega2021experimental}%
  \BibitemOpen
  \bibfield  {author} {\bibinfo {author} {\bibfnamefont {E.~A.}\ \bibnamefont
  {Ortega}}, \bibinfo {author} {\bibfnamefont {K.}~\bibnamefont {Dovzhik}},
  \bibinfo {author} {\bibfnamefont {J.}~\bibnamefont {Fuenzalida}}, \bibinfo
  {author} {\bibfnamefont {S.}~\bibnamefont {Wengerowsky}}, \bibinfo {author}
  {\bibfnamefont {J.~C.}\ \bibnamefont {Alvarado-Zacarias}}, \bibinfo {author}
  {\bibfnamefont {R.~F.}\ \bibnamefont {Shiozaki}}, \bibinfo {author}
  {\bibfnamefont {R.}~\bibnamefont {Amezcua-Correa}}, \bibinfo {author}
  {\bibfnamefont {M.}~\bibnamefont {Bohmann}},\ and\ \bibinfo {author}
  {\bibfnamefont {R.}~\bibnamefont {Ursin}},\ }\bibfield  {title} {\bibinfo
  {title} {Experimental space-division multiplexed polarization-entanglement
  distribution through 12 paths of a multicore fiber},\ }\href
  {https://doi.org/10.1103/PRXQuantum.2.040356} {\bibfield  {journal} {\bibinfo
   {journal} {PRX Quantum}\ }\textbf {\bibinfo {volume} {2}},\ \bibinfo {pages}
  {040356} (\bibinfo {year} {2021})}\BibitemShut {NoStop}%
\bibitem [{\citenamefont {Achatz}\ \emph {et~al.}(2023)\citenamefont {Achatz},
  \citenamefont {Bulla}, \citenamefont {Ecker}, \citenamefont {Ortega},
  \citenamefont {Bartokos}, \citenamefont {Alvarado-Zacarias}, \citenamefont
  {Amezcua-Correa}, \citenamefont {Bohmann}, \citenamefont {Ursin},\ and\
  \citenamefont {Huber}}]{achatz2022simultaneous}%
  \BibitemOpen
  \bibfield  {author} {\bibinfo {author} {\bibfnamefont {L.}~\bibnamefont
  {Achatz}}, \bibinfo {author} {\bibfnamefont {L.}~\bibnamefont {Bulla}},
  \bibinfo {author} {\bibfnamefont {S.}~\bibnamefont {Ecker}}, \bibinfo
  {author} {\bibfnamefont {E.~A.}\ \bibnamefont {Ortega}}, \bibinfo {author}
  {\bibfnamefont {M.}~\bibnamefont {Bartokos}}, \bibinfo {author}
  {\bibfnamefont {J.~C.}\ \bibnamefont {Alvarado-Zacarias}}, \bibinfo {author}
  {\bibfnamefont {R.}~\bibnamefont {Amezcua-Correa}}, \bibinfo {author}
  {\bibfnamefont {M.}~\bibnamefont {Bohmann}}, \bibinfo {author} {\bibfnamefont
  {R.}~\bibnamefont {Ursin}},\ and\ \bibinfo {author} {\bibfnamefont
  {M.}~\bibnamefont {Huber}},\ }\bibfield  {title} {\bibinfo {title}
  {Simultaneous transmission of hyper-entanglement in three degrees of freedom
  through a multicore fiber},\ }\href
  {https://www.nature.com/articles/s41534-023-00700-0#citeas} {\bibfield
  {journal} {\bibinfo  {journal} {npj Quantum Inf.}\ }\textbf {\bibinfo
  {volume} {9}},\ \bibinfo {pages} {45} (\bibinfo {year} {2023})}\BibitemShut
  {NoStop}%
\bibitem [{\citenamefont {Walborn}\ \emph {et~al.}(2010)\citenamefont
  {Walborn}, \citenamefont {Monken}, \citenamefont {Pádua},\ and\
  \citenamefont {{Souto Ribeiro}}}]{WALBORN201087}%
  \BibitemOpen
  \bibfield  {author} {\bibinfo {author} {\bibfnamefont {S.}~\bibnamefont
  {Walborn}}, \bibinfo {author} {\bibfnamefont {C.}~\bibnamefont {Monken}},
  \bibinfo {author} {\bibfnamefont {S.}~\bibnamefont {Pádua}},\ and\ \bibinfo
  {author} {\bibfnamefont {P.}~\bibnamefont {{Souto Ribeiro}}},\ }\bibfield
  {title} {\bibinfo {title} {Spatial correlations in parametric
  down-conversion},\ }\href
  {https://doi.org/https://doi.org/10.1016/j.physrep.2010.06.003} {\bibfield
  {journal} {\bibinfo  {journal} {Phys. Rep.}\ }\textbf {\bibinfo {volume}
  {495}},\ \bibinfo {pages} {87} (\bibinfo {year} {2010})}\BibitemShut
  {NoStop}%
\bibitem [{\citenamefont {Achatz}\ \emph {et~al.}(2022)\citenamefont {Achatz},
  \citenamefont {Ortega}, \citenamefont {Dovzhik}, \citenamefont {Shiozaki},
  \citenamefont {Fuenzalida}, \citenamefont {Wengerowsky}, \citenamefont
  {Bohmann},\ and\ \citenamefont {Ursin}}]{Achatz_2022}%
  \BibitemOpen
  \bibfield  {author} {\bibinfo {author} {\bibfnamefont {L.}~\bibnamefont
  {Achatz}}, \bibinfo {author} {\bibfnamefont {E.~A.}\ \bibnamefont {Ortega}},
  \bibinfo {author} {\bibfnamefont {K.}~\bibnamefont {Dovzhik}}, \bibinfo
  {author} {\bibfnamefont {R.~F.}\ \bibnamefont {Shiozaki}}, \bibinfo {author}
  {\bibfnamefont {J.}~\bibnamefont {Fuenzalida}}, \bibinfo {author}
  {\bibfnamefont {S.}~\bibnamefont {Wengerowsky}}, \bibinfo {author}
  {\bibfnamefont {M.}~\bibnamefont {Bohmann}},\ and\ \bibinfo {author}
  {\bibfnamefont {R.}~\bibnamefont {Ursin}},\ }\bibfield  {title} {\bibinfo
  {title} {Certifying position-momentum entanglement at telecommunication
  wavelengths},\ }\href {https://doi.org/10.1088/1402-4896/ac44b5} {\bibfield
  {journal} {\bibinfo  {journal} {Physica Scripta}\ }\textbf {\bibinfo {volume}
  {97}},\ \bibinfo {pages} {015101} (\bibinfo {year} {2022})}\BibitemShut
  {NoStop}%
\bibitem [{\citenamefont {Takenaga}\ \emph {et~al.}(2011)\citenamefont
  {Takenaga}, \citenamefont {Arakawa}, \citenamefont {Tanigawa}, \citenamefont
  {Guan}, \citenamefont {Matsuo}, \citenamefont {Saitoh},\ and\ \citenamefont
  {Koshiba}}]{trenchMCF2011}%
  \BibitemOpen
  \bibfield  {author} {\bibinfo {author} {\bibfnamefont {K.}~\bibnamefont
  {Takenaga}}, \bibinfo {author} {\bibfnamefont {Y.}~\bibnamefont {Arakawa}},
  \bibinfo {author} {\bibfnamefont {S.}~\bibnamefont {Tanigawa}}, \bibinfo
  {author} {\bibfnamefont {N.}~\bibnamefont {Guan}}, \bibinfo {author}
  {\bibfnamefont {S.}~\bibnamefont {Matsuo}}, \bibinfo {author} {\bibfnamefont
  {K.}~\bibnamefont {Saitoh}},\ and\ \bibinfo {author} {\bibfnamefont
  {M.}~\bibnamefont {Koshiba}},\ }\bibfield  {title} {\bibinfo {title}
  {Reduction of crosstalk by trench-assisted multi-core fiber},\ }in\ \href
  {https://doi.org/10.1364/OFC.2011.OWJ4} {\emph {\bibinfo {booktitle} {2011
  Optical Fiber Communication Conference and Exposition and the National Fiber
  Optic Engineers Conference}}}\ (\bibinfo {year} {2011})\ p.\ \bibinfo {pages}
  {OWJ4}\BibitemShut {NoStop}%
\bibitem [{\citenamefont {Alvarado-Zacarias}\ \emph {et~al.}(2019)\citenamefont
  {Alvarado-Zacarias}, \citenamefont {Antonio-Lopez}, \citenamefont {Habib},
  \citenamefont {Gausmann}, \citenamefont {Wang}, \citenamefont {Cruz-Delgado},
  \citenamefont {Li}, \citenamefont {Sch\"{u}lzgen}, \citenamefont
  {Amezcua-Correa}, \citenamefont {Demontmorillon}, \citenamefont {Sillard},\
  and\ \citenamefont {Amezcua-Correa}}]{MCF-JC}%
  \BibitemOpen
  \bibfield  {author} {\bibinfo {author} {\bibfnamefont {J.~C.}\ \bibnamefont
  {Alvarado-Zacarias}}, \bibinfo {author} {\bibfnamefont {J.~E.}\ \bibnamefont
  {Antonio-Lopez}}, \bibinfo {author} {\bibfnamefont {M.~S.}\ \bibnamefont
  {Habib}}, \bibinfo {author} {\bibfnamefont {S.}~\bibnamefont {Gausmann}},
  \bibinfo {author} {\bibfnamefont {N.}~\bibnamefont {Wang}}, \bibinfo {author}
  {\bibfnamefont {D.}~\bibnamefont {Cruz-Delgado}}, \bibinfo {author}
  {\bibfnamefont {G.}~\bibnamefont {Li}}, \bibinfo {author} {\bibfnamefont
  {A.}~\bibnamefont {Sch\"{u}lzgen}}, \bibinfo {author} {\bibfnamefont
  {A.}~\bibnamefont {Amezcua-Correa}}, \bibinfo {author} {\bibfnamefont
  {L.-A.}\ \bibnamefont {Demontmorillon}}, \bibinfo {author} {\bibfnamefont
  {P.}~\bibnamefont {Sillard}},\ and\ \bibinfo {author} {\bibfnamefont
  {R.}~\bibnamefont {Amezcua-Correa}},\ }\bibfield  {title} {\bibinfo {title}
  {Low-loss 19 core fan-in/fan-out device using reduced-cladding graded index
  fibers},\ }in\ \href {https://doi.org/10.1364/OFC.2019.Th3D.2} {\emph
  {\bibinfo {booktitle} {Optical Fiber Communication Conference (OFC) 2019}}}\
  (\bibinfo  {publisher} {Optical Society of America},\ \bibinfo {year}
  {2019})\ p.\ \bibinfo {pages} {Th3D.2}\BibitemShut {NoStop}%
\bibitem [{\citenamefont {Ortega}\ \emph {et~al.}(2023)\citenamefont {Ortega},
  \citenamefont {Fuenzalida}, \citenamefont {Selimovic}, \citenamefont
  {Dovzhik}, \citenamefont {Achatz}, \citenamefont {Wengerowsky}, \citenamefont
  {Shiozaki}, \citenamefont {Neumann}, \citenamefont {Bohmann},\ and\
  \citenamefont {Ursin}}]{ortega2022spatial}%
  \BibitemOpen
  \bibfield  {author} {\bibinfo {author} {\bibfnamefont {E.~A.}\ \bibnamefont
  {Ortega}}, \bibinfo {author} {\bibfnamefont {J.}~\bibnamefont {Fuenzalida}},
  \bibinfo {author} {\bibfnamefont {M.}~\bibnamefont {Selimovic}}, \bibinfo
  {author} {\bibfnamefont {K.}~\bibnamefont {Dovzhik}}, \bibinfo {author}
  {\bibfnamefont {L.}~\bibnamefont {Achatz}}, \bibinfo {author} {\bibfnamefont
  {S.}~\bibnamefont {Wengerowsky}}, \bibinfo {author} {\bibfnamefont {R.~F.}\
  \bibnamefont {Shiozaki}}, \bibinfo {author} {\bibfnamefont {S.~P.}\
  \bibnamefont {Neumann}}, \bibinfo {author} {\bibfnamefont {M.}~\bibnamefont
  {Bohmann}},\ and\ \bibinfo {author} {\bibfnamefont {R.}~\bibnamefont
  {Ursin}},\ }\bibfield  {title} {\bibinfo {title} {Spatial and spectral
  characterization of photon pairs at telecommunication-wavelength from type-0
  spontaneous parametric downconversion},\ }\href
  {https://doi.org/10.1364/JOSAB.475583} {\bibfield  {journal} {\bibinfo
  {journal} {J. Opt. Soc. Am. B}\ }\textbf {\bibinfo {volume} {40}},\ \bibinfo
  {pages} {165} (\bibinfo {year} {2023})}\BibitemShut {NoStop}%
\bibitem [{\citenamefont {Ma}\ \emph {et~al.}(2007)\citenamefont {Ma},
  \citenamefont {Fung},\ and\ \citenamefont {Lo}}]{QKD_Ma}%
  \BibitemOpen
  \bibfield  {author} {\bibinfo {author} {\bibfnamefont {X.}~\bibnamefont
  {Ma}}, \bibinfo {author} {\bibfnamefont {C.-H.~F.}\ \bibnamefont {Fung}},\
  and\ \bibinfo {author} {\bibfnamefont {H.-K.}\ \bibnamefont {Lo}},\
  }\bibfield  {title} {\bibinfo {title} {Quantum key distribution with
  entangled photon sources},\ }\href
  {https://doi.org/10.1103/PhysRevA.76.012307} {\bibfield  {journal} {\bibinfo
  {journal} {Phys. Rev. A}\ }\textbf {\bibinfo {volume} {76}},\ \bibinfo
  {pages} {012307} (\bibinfo {year} {2007})}\BibitemShut {NoStop}%
\bibitem [{\citenamefont {Neumann}\ \emph {et~al.}(2021)\citenamefont
  {Neumann}, \citenamefont {Scheidl}, \citenamefont {Selimovic}, \citenamefont
  {Pivoluska}, \citenamefont {Liu}, \citenamefont {Bohmann},\ and\
  \citenamefont {Ursin}}]{neumann2021}%
  \BibitemOpen
  \bibfield  {author} {\bibinfo {author} {\bibfnamefont {S.~P.}\ \bibnamefont
  {Neumann}}, \bibinfo {author} {\bibfnamefont {T.}~\bibnamefont {Scheidl}},
  \bibinfo {author} {\bibfnamefont {M.}~\bibnamefont {Selimovic}}, \bibinfo
  {author} {\bibfnamefont {M.}~\bibnamefont {Pivoluska}}, \bibinfo {author}
  {\bibfnamefont {B.}~\bibnamefont {Liu}}, \bibinfo {author} {\bibfnamefont
  {M.}~\bibnamefont {Bohmann}},\ and\ \bibinfo {author} {\bibfnamefont
  {R.}~\bibnamefont {Ursin}},\ }\bibfield  {title} {\bibinfo {title} {Model for
  optimizing quantum key distribution with continuous-wave pumped
  entangled-photon sources},\ }\href
  {https://doi.org/10.1103/PhysRevA.104.022406} {\bibfield  {journal} {\bibinfo
   {journal} {Phys. Rev. A}\ }\textbf {\bibinfo {volume} {104}},\ \bibinfo
  {pages} {022406} (\bibinfo {year} {2021})}\BibitemShut {NoStop}%
\bibitem [{\citenamefont {Elkouss}\ \emph {et~al.}(2009)\citenamefont
  {Elkouss}, \citenamefont {Leverrier}, \citenamefont {Alleaume},\ and\
  \citenamefont {Boutros}}]{Elkouss2009}%
  \BibitemOpen
  \bibfield  {author} {\bibinfo {author} {\bibfnamefont {D.}~\bibnamefont
  {Elkouss}}, \bibinfo {author} {\bibfnamefont {A.}~\bibnamefont {Leverrier}},
  \bibinfo {author} {\bibfnamefont {R.}~\bibnamefont {Alleaume}},\ and\
  \bibinfo {author} {\bibfnamefont {J.~J.}\ \bibnamefont {Boutros}},\
  }\bibfield  {title} {\bibinfo {title} {Efficient reconciliation protocol for
  discrete-variable quantum key distribution},\ }in\ \href
  {https://doi.org/10.1109/ISIT.2009.5205475} {\emph {\bibinfo {booktitle}
  {2009 IEEE International Symposium on Information Theory}}}\ (\bibinfo {year}
  {2009})\ pp.\ \bibinfo {pages} {1879--1883}\BibitemShut {NoStop}%
\bibitem [{\citenamefont {Sasaki}\ \emph {et~al.}(2017)\citenamefont {Sasaki},
  \citenamefont {Takenaga}, \citenamefont {Aikawa}, \citenamefont {Miyamoto},\
  and\ \citenamefont {Morioka}}]{sasaki2017single}%
  \BibitemOpen
  \bibfield  {author} {\bibinfo {author} {\bibfnamefont {Y.}~\bibnamefont
  {Sasaki}}, \bibinfo {author} {\bibfnamefont {K.}~\bibnamefont {Takenaga}},
  \bibinfo {author} {\bibfnamefont {K.}~\bibnamefont {Aikawa}}, \bibinfo
  {author} {\bibfnamefont {Y.}~\bibnamefont {Miyamoto}},\ and\ \bibinfo
  {author} {\bibfnamefont {T.}~\bibnamefont {Morioka}},\ }\bibfield  {title}
  {\bibinfo {title} {Single-mode 37-core fiber with a cladding diameter of 248
  $\mu$m},\ }in\ \href {https://doi.org/10.1364/OFC.2017.Th1H.2} {\emph
  {\bibinfo {booktitle} {2017 Optical Fiber Communications Conference and
  Exhibition (OFC)}}}\ (\bibinfo {organization} {IEEE},\ \bibinfo {year}
  {2017})\ p.\ \bibinfo {pages} {Th1H.2}\BibitemShut {NoStop}%
\bibitem [{\citenamefont {Dietrich}\ \emph {et~al.}(2017)\citenamefont
  {Dietrich}, \citenamefont {Harris}, \citenamefont {Blaicher}, \citenamefont
  {Corrigan}, \citenamefont {Morris}, \citenamefont {Freude}, \citenamefont
  {Quirrenbach},\ and\ \citenamefont {Koos}}]{Dietrich:17}%
  \BibitemOpen
  \bibfield  {author} {\bibinfo {author} {\bibfnamefont {P.-I.}\ \bibnamefont
  {Dietrich}}, \bibinfo {author} {\bibfnamefont {R.~J.}\ \bibnamefont
  {Harris}}, \bibinfo {author} {\bibfnamefont {M.}~\bibnamefont {Blaicher}},
  \bibinfo {author} {\bibfnamefont {M.~K.}\ \bibnamefont {Corrigan}}, \bibinfo
  {author} {\bibfnamefont {T.~J.}\ \bibnamefont {Morris}}, \bibinfo {author}
  {\bibfnamefont {W.}~\bibnamefont {Freude}}, \bibinfo {author} {\bibfnamefont
  {A.}~\bibnamefont {Quirrenbach}},\ and\ \bibinfo {author} {\bibfnamefont
  {C.}~\bibnamefont {Koos}},\ }\bibfield  {title} {\bibinfo {title} {Printed
  freeform lens arrays on multi-core fibers for highly efficient coupling in
  astrophotonic systems},\ }\href {https://doi.org/10.1364/OE.25.018288}
  {\bibfield  {journal} {\bibinfo  {journal} {Opt. Express}\ }\textbf {\bibinfo
  {volume} {25}},\ \bibinfo {pages} {18288} (\bibinfo {year}
  {2017})}\BibitemShut {NoStop}%
\bibitem [{\citenamefont {Kim}\ \emph {et~al.}(2006)\citenamefont {Kim},
  \citenamefont {Fiorentino},\ and\ \citenamefont {Wong}}]{kim2006}%
  \BibitemOpen
  \bibfield  {author} {\bibinfo {author} {\bibfnamefont {T.}~\bibnamefont
  {Kim}}, \bibinfo {author} {\bibfnamefont {M.}~\bibnamefont {Fiorentino}},\
  and\ \bibinfo {author} {\bibfnamefont {F.~N.~C.}\ \bibnamefont {Wong}},\
  }\bibfield  {title} {\bibinfo {title} {Phase-stable source of
  polarization-entangled photons using a polarization sagnac interferometer},\
  }\href {https://doi.org/10.1103/PhysRevA.73.012316} {\bibfield  {journal}
  {\bibinfo  {journal} {Phys. Rev. A}\ }\textbf {\bibinfo {volume} {73}},\
  \bibinfo {pages} {012316} (\bibinfo {year} {2006})}\BibitemShut {NoStop}%
\end{thebibliography}%

\end{document}